\documentclass[a4paper,twoside,prd,nofootinbib,preprintnumbers,twocolumn]{revtex4}
\usepackage{amssymb,amsmath,bm,natbib}
\usepackage{color}
\usepackage{slashed}
\usepackage{graphics}
\usepackage{graphicx}
\usepackage[utf8]{inputenc}
\usepackage{tabularx}
\usepackage[caption=false]{subfig}
\usepackage{hyperref}
\usepackage{url}
\usepackage{dsfont}
\usepackage{float} 
\usepackage{cancel}
\usepackage{ulem}

\newcommand{\eq}[1]{Eq.~\eqref{#1}}

\newcommand{\red}[1]{{\color{red} #1}}

\begin{document}
	\preprint{PSI-PR-22-06, ZU-TH 06/22}
\title{Towards excluding a light $Z^\prime$ explanation of $b\to s\ell^+\ell^-$}

\author{Andreas Crivellin}
	\email{andreas.crivellin@cern.ch}
	\affiliation{Physik-Institut, Universit\"at Z\"urich, Winterthurerstrasse 190, CH--8057 Z\"urich, Switzerland}
	\affiliation{Paul Scherrer Institut, CH--5232 Villigen PSI, Switzerland}
	
		\author{Claudio Andrea Manzari}
	\email{claudioandrea.manzari@physik.uzh.ch}
	\affiliation{Physik-Institut, Universit\"at Z\"urich, Winterthurerstrasse 190, CH--8057 Z\"urich, Switzerland}
	\affiliation{Paul Scherrer Institut, CH--5232 Villigen PSI, Switzerland}
	
	\author{Wolfgang Altmannshofer}
	\email{waltmann@ucsc.edu}
	\affiliation{Santa Cruz Institute for Particle Physics and Department of Physics, \\ University of California, Santa Cruz, CA 95064, USA}

\author{Gianluca Inguglia}
	\email{gianluca.inguglia@oeaw.ac.at}
	\affiliation{Institute of High Energy Physics, 1050, Vienna, Austria}
	
\author{Paul Feichtinger}
	\email{paul.feichtinger@oeaw.ac.at}
	\affiliation{Institute of High Energy Physics, 1050, Vienna, Austria}	
	
	\author{Jorge Martin Camalich}
\email{jcamalich@iac.es}
\affiliation{Instituto de Astrof\'{\i}sica de Canarias, C/ V\'{\i}a L\'actea, s/n E38205 - La Laguna, Tenerife, Spain}
\affiliation{Universidad de La Laguna, Departamento de Astrof\'{\i}sica, La Laguna, Tenerife, Spain}

\begin{abstract}
The discrepancies between $b\to s\ell^+\ell^-$ data and the corresponding Standard Model predictions constitute the most significant hints for new physics (at the TeV scale or below)
currently available. In fact, many scenarios that can account for these anomalies have been proposed in the literature. However, only a single light new physics explanation, i.e.~with a mass below the $B$ meson scale, is possible: a light $Z^\prime$ boson. In this article, we point out that improved limits on $B\to K^{(*)}\nu\nu$, including the experimental sensitivities required for a proper treatment of the necessarily sizeable $Z^\prime$ width, together with the forthcoming Belle~II analyses of $e^+e^-\to\mu^+\mu^-+{\rm invisible}$, can rule out a $Z^\prime$ explanation of $b\to s\ell^+\ell^-$ data with a mass below $\approx4\,$GeV. Importantly, such a light $Z^\prime$ is the only viable single particle solution to the $b\to s\ell^+\ell^-$ anomalies predicting $R(K^{(*)})>0$ in high $q^2$ bins, therefore providing an essential consistency test of data. 
\end{abstract}
\maketitle

\section{Introduction}
\label{intro}

In recent years, multiple hints for the violation of lepton flavour universality (LFU),  which is satisfied by the Standard Model (SM) gauge interactions, have been accumulated (see Refs.~\cite{Crivellin:2021sff,Fischer:2021sqw} for recent reviews). Among them, the discrepancies between $b\to s\ell^+\ell^-$ data and the corresponding SM predictions are statistically most significant (see Refs.~\cite{CAMALICH20221,Albrecht:2021tul,London:2021lfn} for an overview). Combining the current measurements of the LFU ratios $R(K^{(*)})$ one observes that several new physics (NP) scenarios are statistically preferred over the SM hypothesis with significances close to $5\sigma$~\cite{Altmannshofer:2021qrr,Geng:2021nhg,Alguero:2021anc,Hurth:2021nsi,Isidori:2021vtc}. Furthermore, global analyses including also muon-specific observables, like the branching ratio of $B_s\to\phi\mu^+\mu^-$ and angular observables in $B\to K^*\mu^+\mu^-$, like $P_5^\prime$, show  preferences compared to the SM hypothesis with pulls of up to even more than $7\sigma$, depending on theoretical assumptions and data included in the fits~\cite{Altmannshofer:2021qrr,Geng:2021nhg,Alguero:2021anc,Hurth:2021nsi,Kowalska:2019ley,Ciuchini:2021smi,DAmico:2017mtc}.

Because the $b\to s\ell^+\ell^-$ anomalies constitute such tantalizing hints for NP, a plethora of SM extensions have been proposed in the literature, including leptoquarks~\cite{Alonso:2015sja, Calibbi:2015kma,Hiller:2016kry,Bhattacharya:2016mcc, Buttazzo:2017ixm,Barbieri:2015yvd,Barbieri:2016las, Calibbi:2017qbu, Crivellin:2017dsk,Bordone:2018nbg,Kumar:2018kmr, Crivellin:2018yvo,Crivellin:2019szf,Cornella:2019hct, Bordone:2019uzc,Bernigaud:2019bfy,Aebischer:2018acj,Fuentes-Martin:2019ign,Popov:2019tyc,Fajfer:2015ycq,Blanke:2018sro,deMedeirosVarzielas:2019lgb,Varzielas:2015iva,Crivellin:2019dwb,Saad:2020ihm,Saad:2020ucl,Gherardi:2020qhc,DaRold:2020bib,Heeck:2022znj}, models with loop effects of new scalars and fermions~\cite{Gripaios:2015gra,Arnan:2016cpy,Grinstein:2018fgb,Li:2018rax,Marzo:2019ldg,Crivellin:2019dun,Arnan:2019uhr} and in particular models with new neutral gauge bosons, i.e.~$Z^\prime$s~\cite{Buras:2013qja,Gauld:2013qba,Gauld:2013qja,Altmannshofer:2014cfa,Crivellin:2015mga,Crivellin:2015lwa,Niehoff:2015bfa,Sierra:2015fma,Carmona:2015ena,Falkowski:2015zwa,Celis:2015eqs,Celis:2015ara,Crivellin:2015era,Boucenna:2016wpr,Altmannshofer:2016oaq,Boucenna:2016qad,Crivellin:2016ejn,GarciaGarcia:2016nvr,Faisel:2017glo,King:2017anf,Chiang:2017hlj,DiChiara:2017cjq,Ko:2017lzd,Sannino:2017utc,Raby:2017igl,Alonso:2017bff,Cline:2017ihf,Carmona:2017fsn,Falkowski:2018dsl,Benavides:2018rgh,Maji:2018gvz,Singirala:2018mio,Guadagnoli:2018ojc,Allanach:2018lvl,Kohda:2018xbc,King:2018fcg,Duan:2018akc,Rocha-Moran:2018jzu,Dwivedi:2019uqd,Foldenauer:2019vgn,Ko:2019tts,Allanach:2019iiy,Kawamura:2019rth,Altmannshofer:2019xda,Calibbi:2019lvs,Aebischer:2019blw,Kawamura:2019hxp,Crivellin:2020oup,Allanach:2020kss,Capdevila:2020rrl,Greljo:2021xmg,Davighi:2021oel,Allanach:2021kzj,Navarro:2021sfb,Ko:2021lpx,Bause:2021prv,Allanach:2021gmj,Alguero:2022est}. Because all these solutions, except the $Z^\prime$ one, involve charged particles they must be realized at the electroweak scale, or even significantly above, due to the constraints from direct searches. However, a $Z^\prime$ boson can be light and, in fact, such solutions to the $b\to s\ell^+\ell^-$ anomalies have been proposed and studied in the literature~\cite{Sala:2017ihs,Mohapatra:2021izl,Datta:2017ezo,Altmannshofer:2017bsz,Sala:2018ukk,Bishara:2017pje,Borah:2020swo,Darme:2021qzw}. Importantly, this is the only  NP model addressing the $b\to s\ell^+\ell^-$ anomalies which predicts $R(K^{(*)})>0$ in the high $q^2$ bins, i.e.~above the charm resonances, where precise experimental data is still missing but expected in the near future.

While a light $Z^\prime$ explanation of $b\to s\ell^+\ell^-$ data is experimentally well constrained, it still remains viable if it is assumed that the $Z^\prime$ decays dominantly to invisible final states. This avoids direct searches such as $e^+e^-\to 4\mu$ and provides the sizable width necessary for the $Z^\prime$ to affect multiple $q^2$ bins in $b\to s\ell^+\ell^-$ observables and thus give a good fit to data. However, also processes with invisible final states are constrained experimentally, such as the di-muon invariant mass distribution in Drell-Yan production close to the $Z$ mass~\cite{Bishara:2017pje} or $e^+e^-\to \mu^+\mu^-+$invisible at Belle~II~\cite{Adachi:2019otg}. In this letter, we analyze these processes together with $B\to K^{(*)}\nu\bar\nu$~\cite{Belle:2017oht} using a proper treatment of the $Z^\prime$ contribution, including the effects of its large width and the mass of the invisible states it decays to. We assess the possibility that a light $Z^\prime$ can in fact be responsible for the $b\to s\ell^+\ell^-$ anomalies and how this option can be tested in the future with the forthcoming BELLE~II analyses.

\section{Setup and observables}

As outlined in the introduction, we supplement the SM by a light neutral gauge boson, i.e.~a $Z^\prime$, with a mass below the $B$-meson mass scale ($m_{Z^\prime}\lesssim6$~GeV). We will be agnostic about the origin of this new state and simply parametrize its couplings necessary to explain $b\to s\ell^+\ell^-$ data by the Lagrangian
	\begin{eqnarray}
		{\cal L}_{Z^\prime} \supset\left(\bar \mu \left(g_{\mu\mu}^V 
		\gamma^\mu +g_{\mu\mu}^A 
		\gamma^\mu \gamma_5 \right)\mu + 
		g_{sb}^{L,R} \bar s \gamma^\mu P_{L,R} b\right)Z^\prime_\mu\,.
	\end{eqnarray}
We solely include couplings to muons because the ones to electrons are not necessary to explain the $b\to s\ell^+\ell^-$ anomalies and are experimentally well constrained, in particular when they appear simultaneously with muon couplings. Furthermore, we do not consider couplings to muon neutrinos as they are very stringently constrained by neutrino trident production. Note that this is possible even for left-handed muon couplings because we assume our model to be realized below the EW symmetry breaking scale such that $SU(2)_L$ invariance is not necessarily obeyed by the $Z^\prime$ couplings.

Furthermore, in order to achieve the large width necessary to affect multiple $q^2=s$ bins in $b\to s\ell^+\ell^-$ observables such that a good fit to data is possible, we will assume that the $Z^\prime$ has a large decay rate to invisible final states $\chi$ with $m_\chi<m_{Z^\prime}/2$. As the couplings to $\bar sb$ and $\bar \mu\mu$ turn out to be small, we will assume that the branching ratio to invisible final states is to a good approximation 100\%. Furthermore, for specificity $\chi$ is taken to be a fermion with vectorial couplings to the $Z^\prime$.\footnote{This is relevant for calculating the width of the $Z^\prime$ as a function of $q^2$. However, we checked that this assumption has a minor impact on our final results.}

\subsection{$b\to s\ell^+\ell^-$}

Using the standard parametrization of semi-leptonic $B$ decays, the effect of a light $Z^\prime$ can be described by a $q^2$ dependent contribution to the effective Wilson coefficients
\begin{align}
\begin{split}
    C_{9(10)} = \frac{g_{sb}^Lg^{V(A)}_{\mu\mu}}{q^2-m_{Z^{\prime}}^2+i m_{Z^{\prime}}\Gamma_{Z^{\prime}}}\,,\\
    C_{9(10)}^{\prime} = \frac{g_{sb}^Rg^{V(A)}_{\mu\mu}}{q^2-m_{Z^{\prime}}^2+i m_{Z^{\prime}}\Gamma_{Z^{\prime}}}\,,
\end{split}
\end{align}
defined at the $B$ meson scale. Here $\Gamma_{Z^{\prime}}$ is the width of the light vector boson which we approximate here to be $q^2$ independent. For the phenomenological analysis, we implemented these  contributions into \verb|flavio|~\cite{Straub:2018kue} to perform the global fit of our model to data. This includes e.g.~the measurement of the LFU ratios $R_{K}$~\cite{LHCb:2021trn}, $R_{K^{(*)}}$~\cite{LHCb:2017avl}, $R_{K^0_S}$~\cite{LHCb:2021lvy} and $R_{K^{*+}}$~\cite{LHCb:2021lvy}, as well as the branching ratio for $B_s\to\mu^+\mu^-$~\cite{CMS:2014xfa,ATLAS:2018cur,CMS:2019bbr}, the angular observables of $B\to K^{*}\mu^+\mu^-$~\cite{LHCb:2020gog} and the branching ratio and angular distribution of $B_s\to \phi\mu^+\mu^-$~\cite{LHCb:2021xxq,LHCb:2021zwz} which exhibit the most significant deviations from SM predictions.

\subsection{$B\to K^{(*)}+{\rm invisible}$}

The most important constraints on $Z^{\prime}-b-s$ couplings, in case the $Z^\prime$ decays dominantly to invisible final states, can be obtained from the processes $B\to K^{(*)}\nu\bar\nu$ measured most precisely at BaBar~\cite{BaBar:2013npw} and Belle~\cite{Belle:2017oht}. However, only the latest Belle II analysis~\cite{Belle-II:2021rof} with the bound
\begin{equation}
\label{eq:BtoKnunubarBELLE2}
 \mathcal{B}(B^+\to K^+ \nu\bar\nu)<4.1\times10^{-5}\,,
\end{equation}
provides the necessary $q^2$ dependence of the experimental efficiency necessary to easily recast it in terms of the decay $B^+\to K^+\chi\bar\chi$.\footnote{The bounds obtained using the full Belle~\cite{Belle:2017oht} and BaBar~\cite{BaBar:2013npw} data sets are slightly more stringent than Eq.~\eqref{eq:BtoKnunubarBELLE2} but are more difficult to use for our purposes (see, however, Ref.~\cite{MartinCamalich:2020dfe} for a recast of the BaBar limits in terms of a QCD axion).}  

In the case of a large $Z^\prime$ width, the branching ratio $\mathcal{B}(B\to K^{(*)}\chi\bar\chi)$ can be approximated by
\begin{align}
	\begin{split}
		&\mathcal{B}(B\to K^{(*)}\chi\bar{\chi}) = 
		\\ 
		&\qquad  \int_{s_\text{min}}^{s_\text{max}} ds \, \Gamma_{Z^\prime}(s)\, \text{BW}(s) \mathcal{B}(B\to K^{(*)}Z^{\prime})(s)\,,
	\end{split}
	\label{GammaX}
\end{align}
with $s_\text{min}=4m_{\chi}^2$, $s_\text{max}=(m_B-m_{K^{(*)}})^2$ and $\text{BW}(s)=\pi^{-1}\sqrt{s}/((s-m^2_{Z^\prime})^2+\Gamma_
{Z^\prime}(s)^2 m^2_{Z^\prime})$. In these expressions we have kept the $s$-dependence of the $Z^\prime$ width obtained from our fermionic model of the dark sector, $\Gamma_
{Z^\prime}(s)=g_V^2/(12\pi\sqrt{s})(1-4m_\chi^2/s)^{1/2} (s+2m_\chi^2)$, with $g_V$ adjusted such that $\Gamma_
{Z^\prime}(m_{Z^\prime}^2)$ gives the desired width $\Gamma_{Z^\prime}$. The reason for keeping  the $s$-dependence is that it can affect significantly the limits obtained from $B\to K^{(*)}\nu\bar\nu$ searches for large $m_\chi$.

\begin{figure}
	\centering 
	\includegraphics[width=0.45\textwidth]{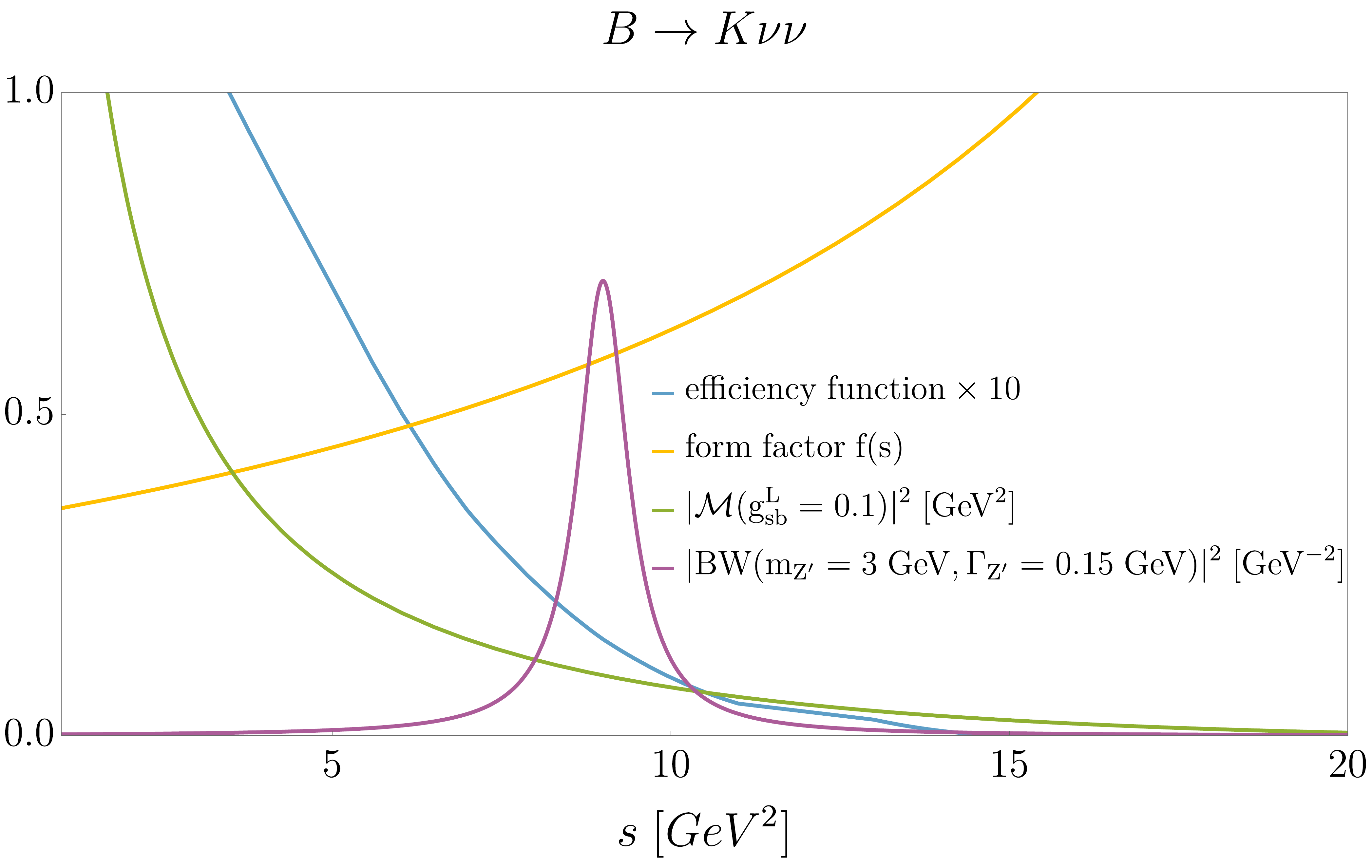}
	\caption{Components of the calculation of ${\mathcal B}(B\to K\nu\bar\nu)$ for a light $Z^\prime$ with a width of 15\%. The final result is obtained by integrating the product of these function starting at $s_{\rm min}=4m_\chi^2$. As the Breit-Wigner leaks to the left where the experimental efficiency and the matrix element (squared) are enhanced, the bounds are stronger for a larger width.}%
	\label{fig:BKnunuDecomposition}%
\end{figure}

With the SM predictions for the differential decay width d$\Gamma(B^+\to K^+\nu\bar\nu)/dq^2$~\cite{Buras:2014fpa}, the relevant form factors~\cite{Bailey:2015dka} and the experimental efficiency function reported by Belle II~\cite{Belle-II:2020jti}, we can translate Eq.~\eqref{eq:BtoKnunubarBELLE2} into a limit on our $Z^\prime$ model, given the masses $m_{Z^\prime}$ and $m_{\chi}$ as well as the width $\Gamma_{Z^\prime}$. 
The experimental signal efficiency~\cite{Belle-II:2020jti} is shown in Fig.~\ref{fig:BKnunuDecomposition} together with the form factor, the Breit-Wigner distribution of the $Z^\prime$ and the squared matrix element of the amplitude (excluding the form factor). The resulting branching ratio is obtained by integrating the product, starting at $s_{\rm min}$. As the amplitude and the efficiency function increase at small $s=q^2$, the bounds on $g_{sb}$ are stronger in case of a large width compared to a narrow one. 

Finally, let us remark that $B\to K\chi\bar\chi$ is only sensitive to the vector current $g_{sb}^L+g_{sb}^R$ such that data from $B\to K^{*}\chi\bar\chi$ would be required to probe the axial-vector coupling $g_{sb}^R-g_{sb}^L$.  However, the former process is sufficient to constraint the NP scenarios needed to explain the $b\to s\ell^+\ell^-$ anomalies as right-handed $bs$ couplings are bounded by the fits  to data.  

\subsection{$B_s-\bar B_s$ mixing}

Tree-level exchange of the $Z^\prime$ contributes to $B_s-\bar B_s$ mixing. For light $Z^\prime$ masses one can set up an operator product expansion in  $m_{Z^\prime}/m_b$ to calculate this new physics contribution to the mixing amplitude and obtain bounds on $g_{sb}$~\cite{MartinCamalich:2020dfe}. However, these limits turn out the be much weaker than the ones from $B\to K^{(*)}\nu\nu$. While in principle for higher $Z^\prime$ masses there could be a (close to) resonant enhancement, it is not clear how to calculate these effects reliably and we will therefore not use $B_s-\bar B_s$ mixing as a constraint in our analysis.

\subsection{$(g-2)_\mu$}

The anomalous magnetic moment of the muon receives 1-loop corrections from the $Z^\prime$. With the results given e.g.~in Ref.~\cite{Buras:2021btx} we find
\begin{align}
\Delta a_{\mu}=&\frac{m_\mu^2}{12\pi^2 M_{Z^{\prime}}^2}{\rm Re}\left[(g_{{\mu}{\mu}}^{V})^2 - 5(g_{{\mu}{\mu}}^{A})^2\right]\,.
\label{AMM}
\end{align}
This expression has to be compared with the experimental value~\cite{Bennett:2006fi,Abi:2021gix} and the SM prediction~\cite{Aoyama:2020ynm}\footnote{This result is based on Refs.~\cite{Aoyama:2012wk,Aoyama:2019ryr,Czarnecki:2002nt,Gnendiger:2013pva,Davier:2017zfy,Keshavarzi:2018mgv,Colangelo:2018mtw,Hoferichter:2019gzf,Davier:2019can,Keshavarzi:2019abf,Kurz:2014wya,Melnikov:2003xd,Masjuan:2017tvw,Colangelo:2017fiz,Hoferichter:2018kwz,Gerardin:2019vio,Bijnens:2019ghy,Colangelo:2019uex,Blum:2019ugy,Colangelo:2014qya}. The recent lattice result of the Budapest-Marseilles-Wuppertal collaboration (BMWc) for the hadronic vacuum polarization (HVP)~\cite{Borsanyi:2020mff}, on the other hand, is not included. This result would render the SM prediction of $a_\mu$ compatible with experiment. However, the BMWc results are in tension with the HVP determined from $e^+e^-\to$ hadrons data~\cite{Davier:2017zfy,Keshavarzi:2018mgv,Colangelo:2018mtw,Hoferichter:2019gzf,Davier:2019can,Keshavarzi:2019abf}. Furthermore, the HVP also enters the global EW fit~\cite{Passera:2008jk}, whose (indirect) determination is below the BMWc result~\cite{Haller:2018nnx}. Therefore, the BMWc determination of the HVP would increase tension in the EW fit~\cite{Crivellin:2020zul,Keshavarzi:2020bfy} and we opted for using the community consensus of Ref.~\cite{Aoyama:2020ynm}.}, resulting in $\Delta a_\mu=a_\mu^{\rm exp} -a_\mu^{\rm SM}=251(59)\times 10^{-11}$.

\subsection{$pp \to \mu^+\mu^-(+{\rm anything})$}

Ref.~\cite{Bishara:2017pje} pointed out that Drell-Yan (DY) searches for muon pairs at the LHC place relevant limits on the parameter space. The $Z^\prime$ can be radiated from the final state muons and significantly modify the di-muon invariant mass distribution close the the $Z$ pole. It is found that for a $Z^\prime$ mass between $1-5\,$GeV the muon coupling should be smaller than $\approx 0.1$ in case of a dominant branching ratio to invisible.

\subsection{$e^+e^- \to \mu^+\mu^-+{\rm invisible}$}
The Belle~II experiment released a search of invisible $Z^\prime$ decays in the process $e^+e^- \to \mu^+\mu^- +$invisible~\cite{Adachi:2019otg} using the commissioning run data. Although limited by the size of the data sample analyzed (276 pb$^{-1}$), 90\% confidence level limits on the coupling $g_{\mu\mu}^V$ of the order of $10^{-2}-10^{-1}$ were obtained. Belle~II has also provided sensitivity projections for this model for integrated luminosities up to 50 fb$^{-1}$~\cite{belle2projections}; in addition, we obtain projections of the sensitivity up to 5 ab$^{-1}$ by accounting for a scaling factor equivalent to L$^{1/4}$.
While Ref.~\cite{Adachi:2019otg} gives bounds on the vectorial coupling, the cross section for $e^+e^-\to \mu^+\mu^-+Z^\prime$ scales as $(g^{V}_{\mu\mu})^2+(g^{A}_{\mu\mu})^2$ and thus can be easily adjusted to the case of other chiralities. Note that the analysis of Ref.~\cite{Adachi:2019otg} was done for a $Z^\prime$ with a narrow width. We therefore recasted the analysis such that it applies to our case with a sizable $Z^\prime$ width by recalculating the expected signal yield in each bin of the original analysis, assuming a Breit-Wigner with $\Gamma_{Z^\prime}=0.1 M_{Z^{\prime}}$ (Fig.\ref{fig:combined} left) and $\Gamma_{Z^\prime}=0.15 M_{Z^{\prime}}$ (Fig.\ref{fig:combined} right) convoluted with a Gaussian resolution function for the signal. We then set up a binned likelihood fit and used the profile likelihood ratio method to extract the 90\% C.L.~intervals.

\section{Phenomenology}
\label{pheno}

\begin{figure}
	\centering 
	\includegraphics[width=0.45\textwidth]{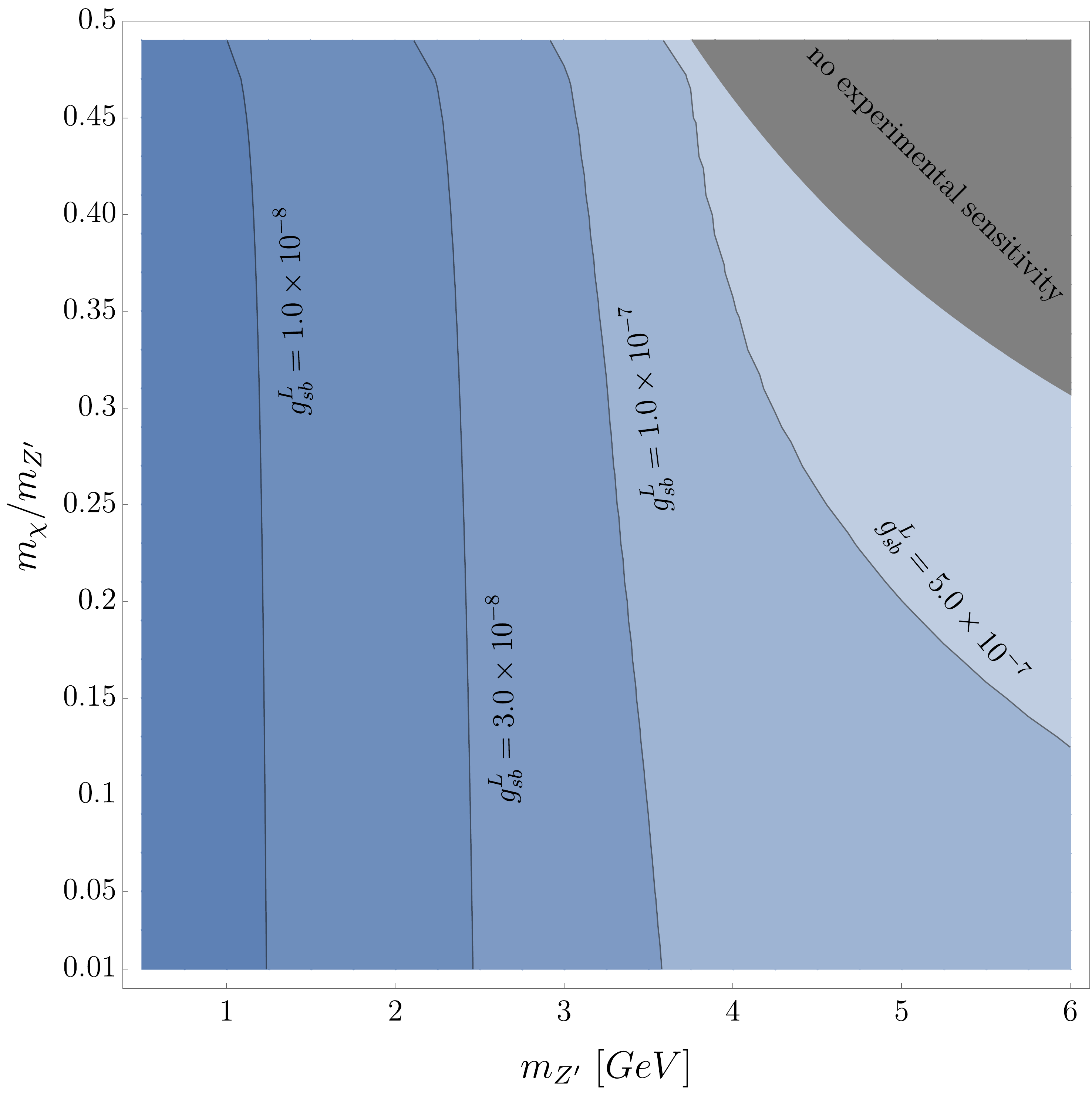}
	\caption{ Contour lines of the bounds on $g_{sb}^L$ in the $m_{Z^\prime}$-$m_\chi/m_{Z^\prime}$ plane for a $Z^\prime$ width of 10\%. The region to the top-right is not constrained as in this case the experimental sensitivity vanishes due $s_{\rm min}=(2m_\chi)^2$.}%
	\label{fig:BKplot}%
\end{figure}

\begin{figure*}[ht!]
	\centering 
	\includegraphics[width=0.49\textwidth]{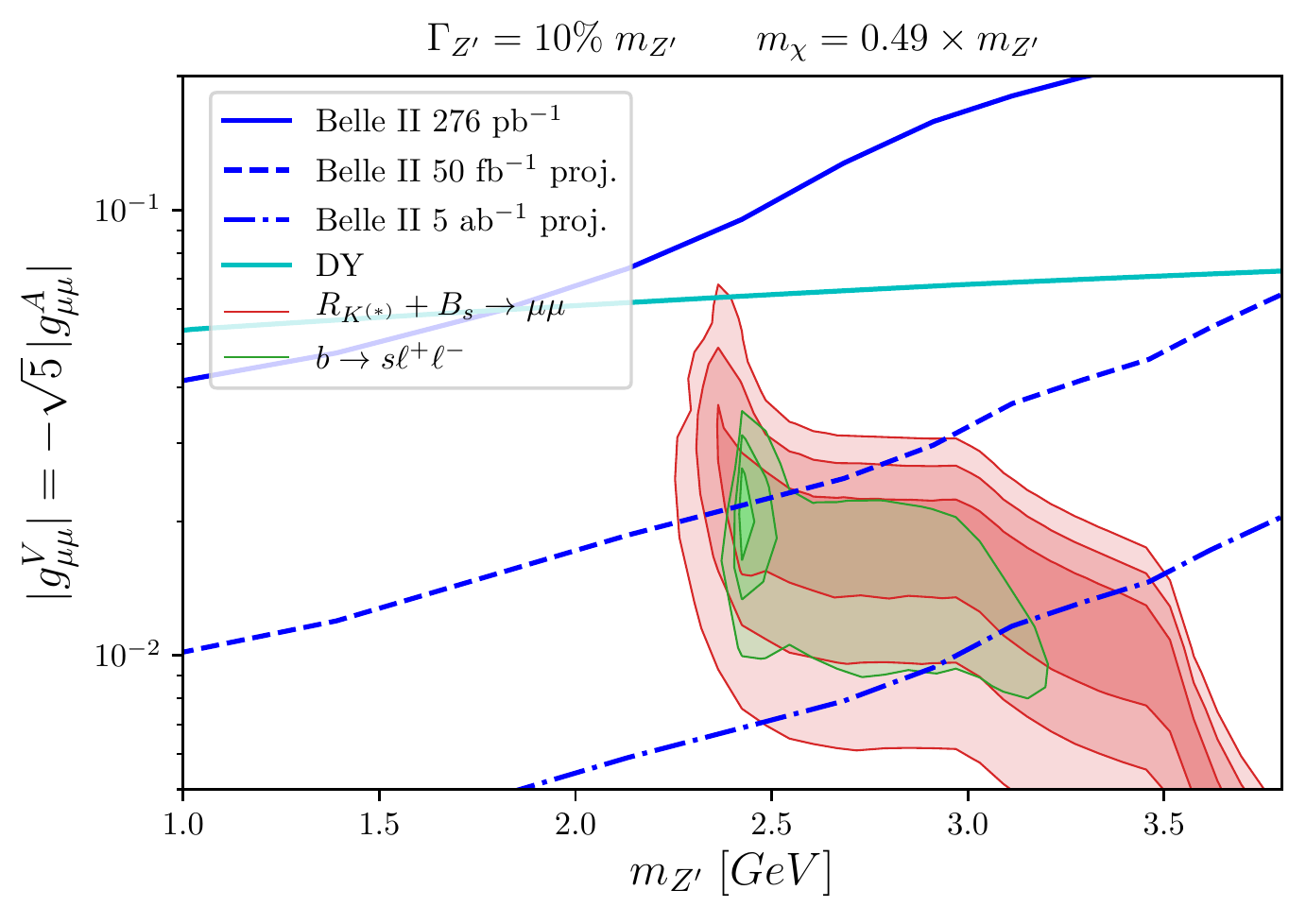}
	\includegraphics[width=0.5\textwidth]{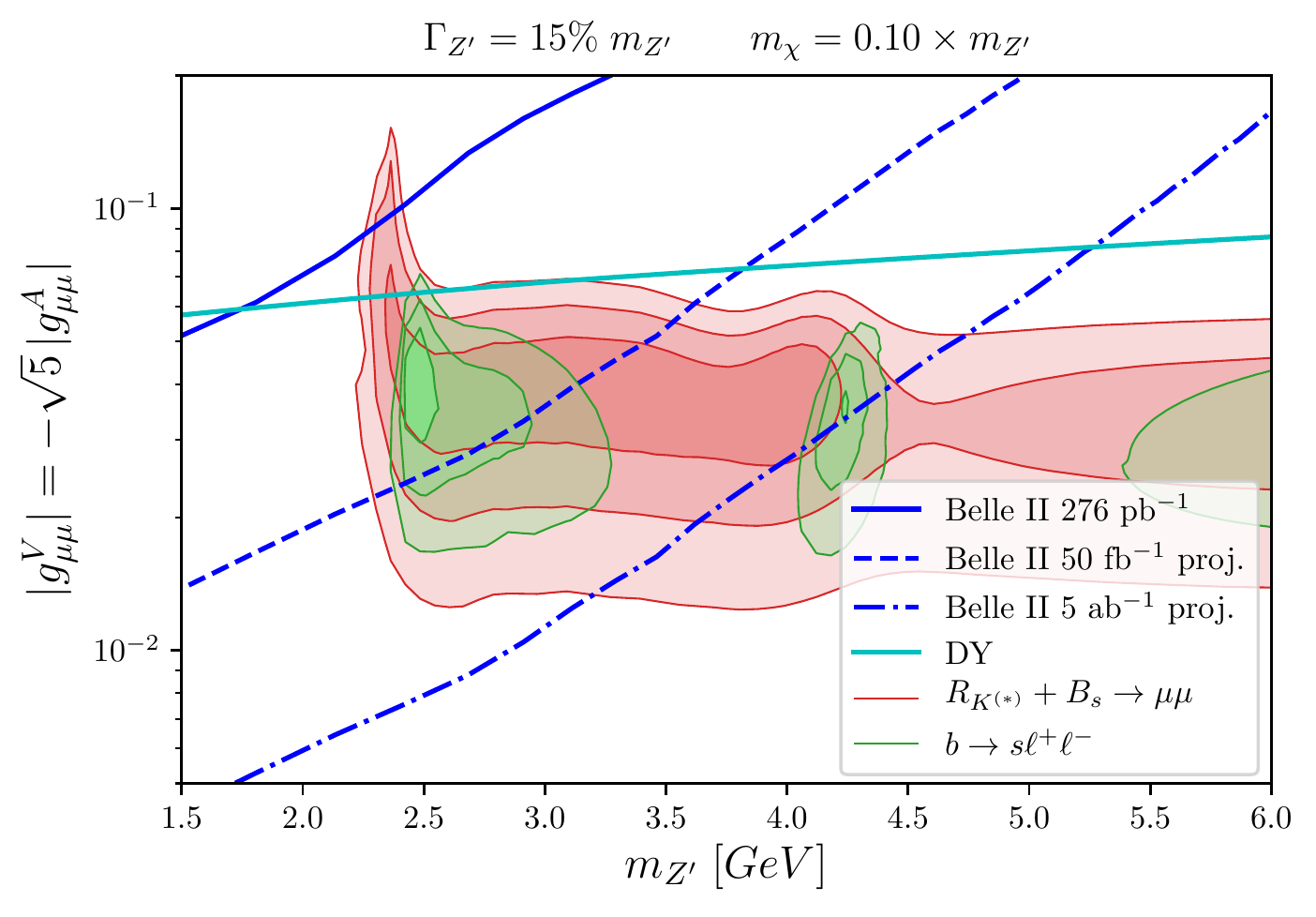}
	\caption{Preferred regions in the $m_{Z^{\prime}}-g^V_{\mu\mu}$ plane from $b\to s\ell^+\ell^-$ (whole data-set, green) and the fit to the LFU ratios $R(K)$ and $B_s\to\mu^+\mu^-$ (red) at the $1\sigma\,, 2\sigma$ and $3\sigma$ level for $g^V_{\mu\mu}=-\sqrt{5}\,g^A_{\mu\mu}$ assuming that $g_{sb}^L$ takes it maximally allowed value from $B\to K\nu\bar\nu$ for different $Z^\prime$ widths and $\chi$ masses. The regions above the solid lines are excluded by the current DY (cyan) and $e^+e^-\to\mu^+\mu^-$invisible searches (blue) while the dashed lines indicate future sensitivities. Note that a smaller width and a larger $\chi$ mass lead to weaker constraints on the model.}
	\label{fig:combined}%
\end{figure*}

First of all, as already noted in Ref.~\cite{Sala:2017ihs}, a sizable width for the $Z^\prime$ is necessary such that it does not only affect a single bin of $P_5^\prime$, $R(K^{(*)})$ etc. This can be achieved by assuming that the $Z^\prime$ decays dominantly into invisible final states $\chi$ which at the same time avoids constraints from searches like $e^+e^-\to 4\mu$. Recasting the $B^+\to K^+\nu\bar\nu$ analysis of Belle~II the limits on $g_{sb}^L$ for a $100\%$ branching ratio to undetected final states\footnote{Of course, the actual branching ratio cannot be 100\% since decays to muons must be possible where kinematically allowed. However, as long as  $Z^\prime\to$invisible is the dominant decay mode, the bounds depend weakly on the branching ratio.} are shown in Fig.~\ref{fig:BKplot}. In this plot we see that a large  $m_\chi\leq m_{Z^\prime}/2$ weakens the bound on $g_{sb}^L$ such that for $2 m_\chi^2\gtrsim15 {\rm GeV}^2$ no limit can be obtained because the experimental sensitivity vanishes. 

Let us now turn to the couplings of the $Z^\prime$ to leptons. For purely vectorial couplings, the bound from $(g-2)_\mu$ would be so strong that it would exclude a $Z^\prime$ explanation of $b\to s\ell^+\ell^-$. However, the effect vanishes for $g_V=-\sqrt{5}g_A$. As this scenario (i.e. $C_9^{\rm eff}=-\sqrt{5}C_{10}^{\rm eff}$) gives a good fit to $b\to s\ell^+\ell^-$ data (as any scenario between the limiting cases $C_9$ and $C_9=-C_{10}$) we will use it as a benchmark scenario here. Note that if we choose $g_V$ slightly bigger, we could account for the tension in $(g-2)_\mu$ while leaving the $b\to s\ell^+\ell^-$ fit unchanged to a very good approximation. 

Performing the $b\to s\ell^+\ell^-$ fit under these assumptions, we have three free parameters, $m_{Z^\prime}$, $\Gamma_{Z^\prime}$ and $g_{sb}^L\times g_{\mu\mu}^V$ (with $g_{\mu\mu}^V=-\sqrt{5}g_{\mu\mu}^A)$. First of all, note that a width $\approx15\%$ gives the best fit to data with $\Delta \chi^2 = \chi^2 - \chi^2_\text{SM} \approx 40$ which is however still smaller than what can be achieved with heavy NP that give a $q^2$ independent effect in the same scenario. Furthermore, $\Delta \chi^2$ does not change significantly for $0.1m_{Z^{\prime}}<\Gamma_{Z^\prime}<0.2m_{Z^{\prime}}$.  

\begin{figure}
	\centering \includegraphics[width=0.5\textwidth]{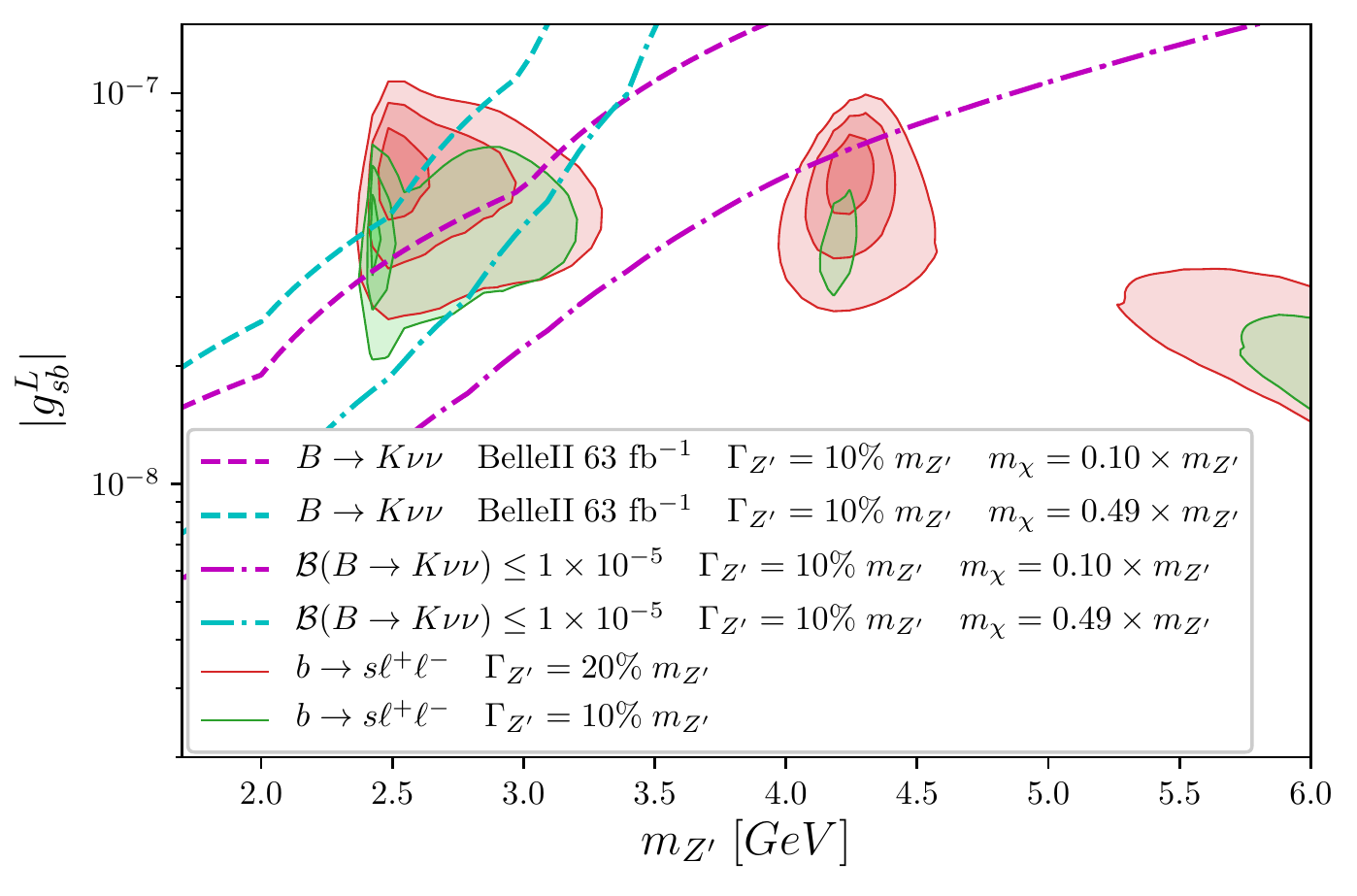}
	\caption{Preferred regions from $b\to s\ell^+\ell^-$ ($1\sigma\,, 2\sigma$ and $3\sigma$) in the $m_{Z^{\prime}}-g^L_{sb}$ plane for the scenario with $g^V_{\mu\mu}=-\sqrt{5}\,g^A_{\mu\mu}$ assuming that $g^V_{\mu\mu}=-\sqrt{5}\,g^A_{\mu\mu}$ takes its maximally allowed value allowed by the $50 {\rm fb}^{-1}$ sensitivity of Belle~II for $e^+e^-\to\mu^+\mu^-$invisible. The regions above the lines, depending on the width and $m_\chi$, can be excluded by future $B\to K\nu\nu$ bounds. }
	\label{fig:combined2}%
\end{figure}

In order to minimize the effect in direct searches for the $Z^\prime$ (i.e.~DY and $e^+e^-\to \mu^+ \mu^- $invisible), given that it provides an explanation to $b\to s\ell^+\ell^-$ data, we can assume that $g^{L}_{sb}$ takes its maximal value allowed by $B^+\to K^+\nu\bar\nu$. The resulting regions preferred by $b\to s\ell^+\ell^-$ data in the $m_{Z^\prime}$ and $g_{\mu\mu}^V=-\sqrt{5}g_{\mu\mu}^A$ plane are shown in Fig.~\ref{fig:combined}. From there, one can see that the constraints from DY searches at the LHC and $e^+e^-\to\mu^+\mu^-+$invisible cannot exclude a light $Z^\prime$ explanation of the $b\to s\ell^+\ell^-$ anomalies, yet. However, the forthcoming Belle~II analysis of $e^+e^-\to\mu^+\mu^-+$invisible has the potential of excluding a mass below $4\,$GeV depending on $m_\chi$ and the width of the $Z^\prime$. Alternatively, we can use the $e^+e^-\to\mu^+\mu^-+$invisible to derive an upper limit on $g_{\mu\mu}^V=-\sqrt{5}g_{\mu\mu}^A$ and show the exclusions from $B^+\to K^+\nu\bar\nu$ in the $m_{Z^\prime}$-$g_{sb}^L$ as depicted in Fig.~\ref{fig:combined2}, were the $50\, \rm{fb}^{-1}$ prospects of Belle~II have been used. Note, that for a width of $15\%$ a $Z^\prime$ with $4\,{\rm GeV}<m_{Z^\prime}<4.5\,$GeV gives a good fit to data and cannot be excluded due to the vanishing experimental sensitivity in $B^+\to K^+\nu\bar\nu$ for a DM mass close to one half of $m_{Z^\prime}$. However, a $Z^\prime$ with such a mass would not lead to $R(K^{(*)})>1$ in the high $q^2$ bins above the $J/\Psi$ resonances.

\section{Conclusions and Outlook}
\label{conclusion}

In this letter we pointed out that a light $Z^\prime$ explanation (with a mass below 4~GeV) of the $b\to s\ell^+\ell^-$ anomalies can be confirmed or disproved by combining the forthcoming Belle~II searches for $e^+e^-\to\mu^+\mu^++$invisible and $B\to K^{(*)}\nu\bar\nu$. Concerning the latter, it is imperative to properly take into account the sizable $Z^\prime$ width and the experimental efficiencies. This endeavour is very important to limit the number of viable models addressing $b\to s\ell^+\ell^-$, in particular in the absence of a signal in direct searches. Furthermore, a light $Z^\prime$ is the only remaining viable NP explanation of $b\to s\ell^+\ell^-$ for which the high $q^2$ bin (above the charm resonances) in e.g.~$R(K^{(*)})$ could lie above unity (assuming that the situation in the low $q^2$ bins remains unchanged). While a light $Z^\prime$ with a mass between $\approx 4-6$ GeV, that enhances the SM amplitude at high $q^2$, cannot be excluded for $m_\chi\approx m_{Z^\prime}/2$ due to the limited experimental sensitivity of the $B\to K^{(*)}\nu\nu$ analysis to low energetic $K^{(*)}$, this gap could be closed in the future, e.g.~with a reliable calculation of $B_s-\bar B_s$ mixing for such $Z^\prime$ masses.  

\acknowledgments{We thank Luc Schnell for checking the heavy NP fit for the $g^V_{\mu\mu}=-\sqrt{5}\,g^A_{\mu\mu}$ scenario. Support by the Swiss National Science Foundation of the Project \textit{$B$ decays and lepton flavour universality violation} (PP00P21\_76884)  is gratefully acknowledged by A.~Crivellin and C.~A.~Manzari. G.~Inguglia and P.~Feichtinger acknowledge support by the European Research Council under the grant agreement No. 947006 - \textit{InterLeptons}. The research of W.~Altmannshofer is supported by the U.S.~Department of Energy grant number DE-SC0010107.  JMC acknowledges Spanish MINECO grant PGC2018-102016-A-I00 and ``Ram\'on y Cajal'' number RYC-2016-20672.}

\appendix

\section{}\label{app:BKZ}

For the decay width of a $B$ meson into $K^{(*)}$ and a vector of invariant mass $s$, the operator product expansion analysis of Refs.~\cite{Bailey:2015dka,Bharucha:2015bzk} gives
\begin{widetext} 
\begin{align}
\mathcal{B}(B\to KZ^{\prime}) =& \frac{(g_{sb}^L+g_{sb}^R)^2f^2(s)}{64\pi \,s\,m_B^3}\lambda(m_B^2,m_K^2,s)^{3/2}\,,\\
\mathcal{B}(B\to K^*Z^{\prime}) =& \frac{(g_{sb}^L+g_{sb}^R)^2V^2(s)}{8\pi m_B^3(m_B+m_{K^*})^2}\lambda(m_B^2,m_{K^*}^2,s)^{3/2} +\frac{(g_{sb}^R-g_{sb}^L)^2A_2^2(s)}{64\pi m_B^3m_{K^*}^2\,s(m_B+m_{K^*})^2}\lambda(m_B^2,m_{K^*}^2,s)^{5/2}\\
&+\frac{(g_{sb}^R-g_{sb}^L)^2A_1^2(s)(m_B^4+m_{K^*}^4+s^2+10\,s\,m_{K^*}^2-2m_B^2(m_{K^*}^2+s))}{64\pi m_B^3m_{K^*}^2\,s}\lambda(m_B^2,m_{K^*}^2,s)^{1/2}\\
&+\frac{(g_{sb}^R-g_{sb}^L)^2A_1(s)A_2(s)(m_B^2-m_{K^*}^2-s)}{32\pi m_B^3m_{K^*}^2\,s}\lambda(m_B^2,m_{K^*}^2,s)^{3/2}\,,
\end{align}
\end{widetext}
where $f(s),\,V(s),\, A_1(s)$ and $A_2(s)$ are the form factor given in Refs.~\cite{Bailey:2015dka,Bharucha:2015bzk} and $\lambda$ is the K\"all\'en function $\lambda(x,y,z) = x^2+y^2+z^2-2xy-2xz-2yz$.

\bibliography{BIB}

\begin{thebibliography}{164}
\expandafter\ifx\csname natexlab\endcsname\relax\def\natexlab#1{#1}\fi
\expandafter\ifx\csname bibnamefont\endcsname\relax
  \def\bibnamefont#1{#1}\fi
\expandafter\ifx\csname bibfnamefont\endcsname\relax
  \def\bibfnamefont#1{#1}\fi
\expandafter\ifx\csname citenamefont\endcsname\relax
  \def\citenamefont#1{#1}\fi
\expandafter\ifx\csname url\endcsname\relax
  \def\url#1{\texttt{#1}}\fi
\expandafter\ifx\csname urlprefix\endcsname\relax\def\urlprefix{URL }\fi
\providecommand{\bibinfo}[2]{#2}
\providecommand{\eprint}[2][]{\url{#2}}

\bibitem[{\citenamefont{Crivellin and Hoferichter}(2021)}]{Crivellin:2021sff}
\bibinfo{author}{\bibfnamefont{A.}~\bibnamefont{Crivellin}} \bibnamefont{and}
  \bibinfo{author}{\bibfnamefont{M.}~\bibnamefont{Hoferichter}},
  \bibinfo{journal}{Science} \textbf{\bibinfo{volume}{374}},
  \bibinfo{pages}{1051} (\bibinfo{year}{2021}), \eprint{2111.12739}.

\bibitem[{\citenamefont{Fischer et~al.}(2021)}]{Fischer:2021sqw}
\bibinfo{author}{\bibfnamefont{O.}~\bibnamefont{Fischer}} \bibnamefont{et~al.}
  (\bibinfo{year}{2021}), \eprint{2109.06065}.

\bibitem[{\citenamefont{Camalich and Patel}(2022)}]{CAMALICH20221}
\bibinfo{author}{\bibfnamefont{J.~M.} \bibnamefont{Camalich}} \bibnamefont{and}
  \bibinfo{author}{\bibfnamefont{M.}~\bibnamefont{Patel}},
  \bibinfo{journal}{Science Bulletin} \textbf{\bibinfo{volume}{67}},
  \bibinfo{pages}{1} (\bibinfo{year}{2022}), ISSN \bibinfo{issn}{2095-9273},
  \urlprefix\url{https://www.sciencedirect.com/science/article/pii/S2095927321006307}.

\bibitem[{\citenamefont{Albrecht et~al.}(2021)\citenamefont{Albrecht, van Dyk,
  and Langenbruch}}]{Albrecht:2021tul}
\bibinfo{author}{\bibfnamefont{J.}~\bibnamefont{Albrecht}},
  \bibinfo{author}{\bibfnamefont{D.}~\bibnamefont{van Dyk}}, \bibnamefont{and}
  \bibinfo{author}{\bibfnamefont{C.}~\bibnamefont{Langenbruch}},
  \bibinfo{journal}{Prog. Part. Nucl. Phys.} \textbf{\bibinfo{volume}{120}},
  \bibinfo{pages}{103885} (\bibinfo{year}{2021}), \eprint{2107.04822}.

\bibitem[{\citenamefont{London and Matias}(2021)}]{London:2021lfn}
\bibinfo{author}{\bibfnamefont{D.}~\bibnamefont{London}} \bibnamefont{and}
  \bibinfo{author}{\bibfnamefont{J.}~\bibnamefont{Matias}}
  (\bibinfo{year}{2021}), \eprint{2110.13270}.

\bibitem[{\citenamefont{Altmannshofer and
  Stangl}(2021)}]{Altmannshofer:2021qrr}
\bibinfo{author}{\bibfnamefont{W.}~\bibnamefont{Altmannshofer}}
  \bibnamefont{and} \bibinfo{author}{\bibfnamefont{P.}~\bibnamefont{Stangl}},
  \bibinfo{journal}{Eur. Phys. J. C} \textbf{\bibinfo{volume}{81}},
  \bibinfo{pages}{952} (\bibinfo{year}{2021}), \eprint{2103.13370}.

\bibitem[{\citenamefont{Geng et~al.}(2021)\citenamefont{Geng, Grinstein,
  J\"ager, Li, Martin~Camalich, and Shi}}]{Geng:2021nhg}
\bibinfo{author}{\bibfnamefont{L.-S.} \bibnamefont{Geng}},
  \bibinfo{author}{\bibfnamefont{B.}~\bibnamefont{Grinstein}},
  \bibinfo{author}{\bibfnamefont{S.}~\bibnamefont{J\"ager}},
  \bibinfo{author}{\bibfnamefont{S.-Y.} \bibnamefont{Li}},
  \bibinfo{author}{\bibfnamefont{J.}~\bibnamefont{Martin~Camalich}},
  \bibnamefont{and} \bibinfo{author}{\bibfnamefont{R.-X.} \bibnamefont{Shi}},
  \bibinfo{journal}{Phys. Rev. D} \textbf{\bibinfo{volume}{104}},
  \bibinfo{pages}{035029} (\bibinfo{year}{2021}), \eprint{2103.12738}.

\bibitem[{\citenamefont{Alguer\'o et~al.}(2021)\citenamefont{Alguer\'o,
  Capdevila, Descotes-Genon, Matias, and Novoa-Brunet}}]{Alguero:2021anc}
\bibinfo{author}{\bibfnamefont{M.}~\bibnamefont{Alguer\'o}},
  \bibinfo{author}{\bibfnamefont{B.}~\bibnamefont{Capdevila}},
  \bibinfo{author}{\bibfnamefont{S.}~\bibnamefont{Descotes-Genon}},
  \bibinfo{author}{\bibfnamefont{J.}~\bibnamefont{Matias}}, \bibnamefont{and}
  \bibinfo{author}{\bibfnamefont{M.}~\bibnamefont{Novoa-Brunet}}
  (\bibinfo{year}{2021}), \eprint{2104.08921}.

\bibitem[{\citenamefont{Hurth et~al.}(2022)\citenamefont{Hurth, Mahmoudi,
  Santos, and Neshatpour}}]{Hurth:2021nsi}
\bibinfo{author}{\bibfnamefont{T.}~\bibnamefont{Hurth}},
  \bibinfo{author}{\bibfnamefont{F.}~\bibnamefont{Mahmoudi}},
  \bibinfo{author}{\bibfnamefont{D.~M.} \bibnamefont{Santos}},
  \bibnamefont{and}
  \bibinfo{author}{\bibfnamefont{S.}~\bibnamefont{Neshatpour}},
  \bibinfo{journal}{Phys. Lett. B} \textbf{\bibinfo{volume}{824}},
  \bibinfo{pages}{136838} (\bibinfo{year}{2022}), \eprint{2104.10058}.

\bibitem[{\citenamefont{Isidori et~al.}(2021)\citenamefont{Isidori, Lancierini,
  Owen, and Serra}}]{Isidori:2021vtc}
\bibinfo{author}{\bibfnamefont{G.}~\bibnamefont{Isidori}},
  \bibinfo{author}{\bibfnamefont{D.}~\bibnamefont{Lancierini}},
  \bibinfo{author}{\bibfnamefont{P.}~\bibnamefont{Owen}}, \bibnamefont{and}
  \bibinfo{author}{\bibfnamefont{N.}~\bibnamefont{Serra}},
  \bibinfo{journal}{Phys. Lett. B} \textbf{\bibinfo{volume}{822}},
  \bibinfo{pages}{136644} (\bibinfo{year}{2021}), \eprint{2104.05631}.

\bibitem[{\citenamefont{Kowalska et~al.}(2019)\citenamefont{Kowalska, Kumar,
  and Sessolo}}]{Kowalska:2019ley}
\bibinfo{author}{\bibfnamefont{K.}~\bibnamefont{Kowalska}},
  \bibinfo{author}{\bibfnamefont{D.}~\bibnamefont{Kumar}}, \bibnamefont{and}
  \bibinfo{author}{\bibfnamefont{E.~M.} \bibnamefont{Sessolo}},
  \bibinfo{journal}{Eur. Phys. J. C} \textbf{\bibinfo{volume}{79}},
  \bibinfo{pages}{840} (\bibinfo{year}{2019}), \eprint{1903.10932}.

\bibitem[{\citenamefont{Ciuchini et~al.}(2021)\citenamefont{Ciuchini, Fedele,
  Franco, Paul, Silvestrini, and Valli}}]{Ciuchini:2021smi}
\bibinfo{author}{\bibfnamefont{M.}~\bibnamefont{Ciuchini}},
  \bibinfo{author}{\bibfnamefont{M.}~\bibnamefont{Fedele}},
  \bibinfo{author}{\bibfnamefont{E.}~\bibnamefont{Franco}},
  \bibinfo{author}{\bibfnamefont{A.}~\bibnamefont{Paul}},
  \bibinfo{author}{\bibfnamefont{L.}~\bibnamefont{Silvestrini}},
  \bibnamefont{and} \bibinfo{author}{\bibfnamefont{M.}~\bibnamefont{Valli}}
  (\bibinfo{year}{2021}), \eprint{2110.10126}.

\bibitem[{\citenamefont{D'Amico et~al.}(2017)\citenamefont{D'Amico, Nardecchia,
  Panci, Sannino, Strumia, Torre, and Urbano}}]{DAmico:2017mtc}
\bibinfo{author}{\bibfnamefont{G.}~\bibnamefont{D'Amico}},
  \bibinfo{author}{\bibfnamefont{M.}~\bibnamefont{Nardecchia}},
  \bibinfo{author}{\bibfnamefont{P.}~\bibnamefont{Panci}},
  \bibinfo{author}{\bibfnamefont{F.}~\bibnamefont{Sannino}},
  \bibinfo{author}{\bibfnamefont{A.}~\bibnamefont{Strumia}},
  \bibinfo{author}{\bibfnamefont{R.}~\bibnamefont{Torre}}, \bibnamefont{and}
  \bibinfo{author}{\bibfnamefont{A.}~\bibnamefont{Urbano}},
  \bibinfo{journal}{JHEP} \textbf{\bibinfo{volume}{09}}, \bibinfo{pages}{010}
  (\bibinfo{year}{2017}), \eprint{1704.05438}.

\bibitem[{\citenamefont{Alonso et~al.}(2015)\citenamefont{Alonso, Grinstein,
  and Martin~Camalich}}]{Alonso:2015sja}
\bibinfo{author}{\bibfnamefont{R.}~\bibnamefont{Alonso}},
  \bibinfo{author}{\bibfnamefont{B.}~\bibnamefont{Grinstein}},
  \bibnamefont{and}
  \bibinfo{author}{\bibfnamefont{J.}~\bibnamefont{Martin~Camalich}},
  \bibinfo{journal}{JHEP} \textbf{\bibinfo{volume}{10}}, \bibinfo{pages}{184}
  (\bibinfo{year}{2015}), \eprint{1505.05164}.

\bibitem[{\citenamefont{Calibbi et~al.}(2015)\citenamefont{Calibbi, Crivellin,
  and Ota}}]{Calibbi:2015kma}
\bibinfo{author}{\bibfnamefont{L.}~\bibnamefont{Calibbi}},
  \bibinfo{author}{\bibfnamefont{A.}~\bibnamefont{Crivellin}},
  \bibnamefont{and} \bibinfo{author}{\bibfnamefont{T.}~\bibnamefont{Ota}},
  \bibinfo{journal}{Phys. Rev. Lett.} \textbf{\bibinfo{volume}{115}},
  \bibinfo{pages}{181801} (\bibinfo{year}{2015}), \eprint{1506.02661}.

\bibitem[{\citenamefont{Hiller et~al.}(2016)\citenamefont{Hiller, Loose, and
  Sch\"onwald}}]{Hiller:2016kry}
\bibinfo{author}{\bibfnamefont{G.}~\bibnamefont{Hiller}},
  \bibinfo{author}{\bibfnamefont{D.}~\bibnamefont{Loose}}, \bibnamefont{and}
  \bibinfo{author}{\bibfnamefont{K.}~\bibnamefont{Sch\"onwald}},
  \bibinfo{journal}{JHEP} \textbf{\bibinfo{volume}{12}}, \bibinfo{pages}{027}
  (\bibinfo{year}{2016}), \eprint{1609.08895}.

\bibitem[{\citenamefont{Bhattacharya et~al.}(2017)\citenamefont{Bhattacharya,
  Datta, Gu\'evin, London, and Watanabe}}]{Bhattacharya:2016mcc}
\bibinfo{author}{\bibfnamefont{B.}~\bibnamefont{Bhattacharya}},
  \bibinfo{author}{\bibfnamefont{A.}~\bibnamefont{Datta}},
  \bibinfo{author}{\bibfnamefont{J.-P.} \bibnamefont{Gu\'evin}},
  \bibinfo{author}{\bibfnamefont{D.}~\bibnamefont{London}}, \bibnamefont{and}
  \bibinfo{author}{\bibfnamefont{R.}~\bibnamefont{Watanabe}},
  \bibinfo{journal}{JHEP} \textbf{\bibinfo{volume}{01}}, \bibinfo{pages}{015}
  (\bibinfo{year}{2017}), \eprint{1609.09078}.

\bibitem[{\citenamefont{Buttazzo et~al.}(2017)\citenamefont{Buttazzo, Greljo,
  Isidori, and Marzocca}}]{Buttazzo:2017ixm}
\bibinfo{author}{\bibfnamefont{D.}~\bibnamefont{Buttazzo}},
  \bibinfo{author}{\bibfnamefont{A.}~\bibnamefont{Greljo}},
  \bibinfo{author}{\bibfnamefont{G.}~\bibnamefont{Isidori}}, \bibnamefont{and}
  \bibinfo{author}{\bibfnamefont{D.}~\bibnamefont{Marzocca}},
  \bibinfo{journal}{JHEP} \textbf{\bibinfo{volume}{11}}, \bibinfo{pages}{044}
  (\bibinfo{year}{2017}), \eprint{1706.07808}.

\bibitem[{\citenamefont{Barbieri et~al.}(2016)\citenamefont{Barbieri, Isidori,
  Pattori, and Senia}}]{Barbieri:2015yvd}
\bibinfo{author}{\bibfnamefont{R.}~\bibnamefont{Barbieri}},
  \bibinfo{author}{\bibfnamefont{G.}~\bibnamefont{Isidori}},
  \bibinfo{author}{\bibfnamefont{A.}~\bibnamefont{Pattori}}, \bibnamefont{and}
  \bibinfo{author}{\bibfnamefont{F.}~\bibnamefont{Senia}},
  \bibinfo{journal}{Eur. Phys. J. C} \textbf{\bibinfo{volume}{76}},
  \bibinfo{pages}{67} (\bibinfo{year}{2016}), \eprint{1512.01560}.

\bibitem[{\citenamefont{Barbieri et~al.}(2017)\citenamefont{Barbieri, Murphy,
  and Senia}}]{Barbieri:2016las}
\bibinfo{author}{\bibfnamefont{R.}~\bibnamefont{Barbieri}},
  \bibinfo{author}{\bibfnamefont{C.~W.} \bibnamefont{Murphy}},
  \bibnamefont{and} \bibinfo{author}{\bibfnamefont{F.}~\bibnamefont{Senia}},
  \bibinfo{journal}{Eur. Phys. J. C} \textbf{\bibinfo{volume}{77}},
  \bibinfo{pages}{8} (\bibinfo{year}{2017}), \eprint{1611.04930}.

\bibitem[{\citenamefont{Calibbi et~al.}(2018)\citenamefont{Calibbi, Crivellin,
  and Li}}]{Calibbi:2017qbu}
\bibinfo{author}{\bibfnamefont{L.}~\bibnamefont{Calibbi}},
  \bibinfo{author}{\bibfnamefont{A.}~\bibnamefont{Crivellin}},
  \bibnamefont{and} \bibinfo{author}{\bibfnamefont{T.}~\bibnamefont{Li}},
  \bibinfo{journal}{Phys. Rev. D} \textbf{\bibinfo{volume}{98}},
  \bibinfo{pages}{115002} (\bibinfo{year}{2018}), \eprint{1709.00692}.

\bibitem[{\citenamefont{Crivellin et~al.}(2018)\citenamefont{Crivellin,
  M\"uller, Signer, and Ulrich}}]{Crivellin:2017dsk}
\bibinfo{author}{\bibfnamefont{A.}~\bibnamefont{Crivellin}},
  \bibinfo{author}{\bibfnamefont{D.}~\bibnamefont{M\"uller}},
  \bibinfo{author}{\bibfnamefont{A.}~\bibnamefont{Signer}}, \bibnamefont{and}
  \bibinfo{author}{\bibfnamefont{Y.}~\bibnamefont{Ulrich}},
  \bibinfo{journal}{Phys. Rev. D} \textbf{\bibinfo{volume}{97}},
  \bibinfo{pages}{015019} (\bibinfo{year}{2018}), \eprint{1706.08511}.

\bibitem[{\citenamefont{Bordone et~al.}(2018)\citenamefont{Bordone, Cornella,
  Fuentes-Mart\'\i{}n, and Isidori}}]{Bordone:2018nbg}
\bibinfo{author}{\bibfnamefont{M.}~\bibnamefont{Bordone}},
  \bibinfo{author}{\bibfnamefont{C.}~\bibnamefont{Cornella}},
  \bibinfo{author}{\bibfnamefont{J.}~\bibnamefont{Fuentes-Mart\'\i{}n}},
  \bibnamefont{and} \bibinfo{author}{\bibfnamefont{G.}~\bibnamefont{Isidori}},
  \bibinfo{journal}{JHEP} \textbf{\bibinfo{volume}{10}}, \bibinfo{pages}{148}
  (\bibinfo{year}{2018}), \eprint{1805.09328}.

\bibitem[{\citenamefont{Kumar et~al.}(2019)\citenamefont{Kumar, London, and
  Watanabe}}]{Kumar:2018kmr}
\bibinfo{author}{\bibfnamefont{J.}~\bibnamefont{Kumar}},
  \bibinfo{author}{\bibfnamefont{D.}~\bibnamefont{London}}, \bibnamefont{and}
  \bibinfo{author}{\bibfnamefont{R.}~\bibnamefont{Watanabe}},
  \bibinfo{journal}{Phys. Rev. D} \textbf{\bibinfo{volume}{99}},
  \bibinfo{pages}{015007} (\bibinfo{year}{2019}), \eprint{1806.07403}.

\bibitem[{\citenamefont{Crivellin
  et~al.}(2019{\natexlab{a}})\citenamefont{Crivellin, Greub, M\"uller, and
  Saturnino}}]{Crivellin:2018yvo}
\bibinfo{author}{\bibfnamefont{A.}~\bibnamefont{Crivellin}},
  \bibinfo{author}{\bibfnamefont{C.}~\bibnamefont{Greub}},
  \bibinfo{author}{\bibfnamefont{D.}~\bibnamefont{M\"uller}}, \bibnamefont{and}
  \bibinfo{author}{\bibfnamefont{F.}~\bibnamefont{Saturnino}},
  \bibinfo{journal}{Phys. Rev. Lett.} \textbf{\bibinfo{volume}{122}},
  \bibinfo{pages}{011805} (\bibinfo{year}{2019}{\natexlab{a}}),
  \eprint{1807.02068}.

\bibitem[{\citenamefont{Crivellin and Saturnino}(2019)}]{Crivellin:2019szf}
\bibinfo{author}{\bibfnamefont{A.}~\bibnamefont{Crivellin}} \bibnamefont{and}
  \bibinfo{author}{\bibfnamefont{F.}~\bibnamefont{Saturnino}},
  \bibinfo{journal}{PoS} \textbf{\bibinfo{volume}{DIS2019}},
  \bibinfo{pages}{163} (\bibinfo{year}{2019}), \eprint{1906.01222}.

\bibitem[{\citenamefont{Cornella et~al.}(2019)\citenamefont{Cornella,
  Fuentes-Martin, and Isidori}}]{Cornella:2019hct}
\bibinfo{author}{\bibfnamefont{C.}~\bibnamefont{Cornella}},
  \bibinfo{author}{\bibfnamefont{J.}~\bibnamefont{Fuentes-Martin}},
  \bibnamefont{and} \bibinfo{author}{\bibfnamefont{G.}~\bibnamefont{Isidori}},
  \bibinfo{journal}{JHEP} \textbf{\bibinfo{volume}{07}}, \bibinfo{pages}{168}
  (\bibinfo{year}{2019}), \eprint{1903.11517}.

\bibitem[{\citenamefont{Bordone et~al.}(2020)\citenamefont{Bordone, Cat\`a, and
  Feldmann}}]{Bordone:2019uzc}
\bibinfo{author}{\bibfnamefont{M.}~\bibnamefont{Bordone}},
  \bibinfo{author}{\bibfnamefont{O.}~\bibnamefont{Cat\`a}}, \bibnamefont{and}
  \bibinfo{author}{\bibfnamefont{T.}~\bibnamefont{Feldmann}},
  \bibinfo{journal}{JHEP} \textbf{\bibinfo{volume}{01}}, \bibinfo{pages}{067}
  (\bibinfo{year}{2020}), \eprint{1910.02641}.

\bibitem[{\citenamefont{Bernigaud et~al.}(2020)\citenamefont{Bernigaud,
  de~Medeiros~Varzielas, and Talbert}}]{Bernigaud:2019bfy}
\bibinfo{author}{\bibfnamefont{J.}~\bibnamefont{Bernigaud}},
  \bibinfo{author}{\bibfnamefont{I.}~\bibnamefont{de~Medeiros~Varzielas}},
  \bibnamefont{and} \bibinfo{author}{\bibfnamefont{J.}~\bibnamefont{Talbert}},
  \bibinfo{journal}{JHEP} \textbf{\bibinfo{volume}{01}}, \bibinfo{pages}{194}
  (\bibinfo{year}{2020}), \eprint{1906.11270}.

\bibitem[{\citenamefont{Aebischer et~al.}(2019)\citenamefont{Aebischer,
  Crivellin, and Greub}}]{Aebischer:2018acj}
\bibinfo{author}{\bibfnamefont{J.}~\bibnamefont{Aebischer}},
  \bibinfo{author}{\bibfnamefont{A.}~\bibnamefont{Crivellin}},
  \bibnamefont{and} \bibinfo{author}{\bibfnamefont{C.}~\bibnamefont{Greub}},
  \bibinfo{journal}{Phys. Rev. D} \textbf{\bibinfo{volume}{99}},
  \bibinfo{pages}{055002} (\bibinfo{year}{2019}), \eprint{1811.08907}.

\bibitem[{\citenamefont{Fuentes-Mart\'\i{}n
  et~al.}(2020)\citenamefont{Fuentes-Mart\'\i{}n, Isidori, K\"onig, and
  Selimovi\'c}}]{Fuentes-Martin:2019ign}
\bibinfo{author}{\bibfnamefont{J.}~\bibnamefont{Fuentes-Mart\'\i{}n}},
  \bibinfo{author}{\bibfnamefont{G.}~\bibnamefont{Isidori}},
  \bibinfo{author}{\bibfnamefont{M.}~\bibnamefont{K\"onig}}, \bibnamefont{and}
  \bibinfo{author}{\bibfnamefont{N.}~\bibnamefont{Selimovi\'c}},
  \bibinfo{journal}{Phys. Rev. D} \textbf{\bibinfo{volume}{101}},
  \bibinfo{pages}{035024} (\bibinfo{year}{2020}), \eprint{1910.13474}.

\bibitem[{\citenamefont{Popov et~al.}(2019)\citenamefont{Popov, Schmidt, and
  White}}]{Popov:2019tyc}
\bibinfo{author}{\bibfnamefont{O.}~\bibnamefont{Popov}},
  \bibinfo{author}{\bibfnamefont{M.~A.} \bibnamefont{Schmidt}},
  \bibnamefont{and} \bibinfo{author}{\bibfnamefont{G.}~\bibnamefont{White}},
  \bibinfo{journal}{Phys. Rev. D} \textbf{\bibinfo{volume}{100}},
  \bibinfo{pages}{035028} (\bibinfo{year}{2019}), \eprint{1905.06339}.

\bibitem[{\citenamefont{Fajfer and Ko\v{s}nik}(2016)}]{Fajfer:2015ycq}
\bibinfo{author}{\bibfnamefont{S.}~\bibnamefont{Fajfer}} \bibnamefont{and}
  \bibinfo{author}{\bibfnamefont{N.}~\bibnamefont{Ko\v{s}nik}},
  \bibinfo{journal}{Phys. Lett. B} \textbf{\bibinfo{volume}{755}},
  \bibinfo{pages}{270} (\bibinfo{year}{2016}), \eprint{1511.06024}.

\bibitem[{\citenamefont{Blanke and Crivellin}(2018)}]{Blanke:2018sro}
\bibinfo{author}{\bibfnamefont{M.}~\bibnamefont{Blanke}} \bibnamefont{and}
  \bibinfo{author}{\bibfnamefont{A.}~\bibnamefont{Crivellin}},
  \bibinfo{journal}{Phys. Rev. Lett.} \textbf{\bibinfo{volume}{121}},
  \bibinfo{pages}{011801} (\bibinfo{year}{2018}), \eprint{1801.07256}.

\bibitem[{\citenamefont{de~Medeiros~Varzielas and
  Talbert}(2019)}]{deMedeirosVarzielas:2019lgb}
\bibinfo{author}{\bibfnamefont{I.}~\bibnamefont{de~Medeiros~Varzielas}}
  \bibnamefont{and} \bibinfo{author}{\bibfnamefont{J.}~\bibnamefont{Talbert}},
  \bibinfo{journal}{Eur. Phys. J. C} \textbf{\bibinfo{volume}{79}},
  \bibinfo{pages}{536} (\bibinfo{year}{2019}), \eprint{1901.10484}.

\bibitem[{\citenamefont{de~Medeiros~Varzielas and
  Hiller}(2015)}]{Varzielas:2015iva}
\bibinfo{author}{\bibfnamefont{I.}~\bibnamefont{de~Medeiros~Varzielas}}
  \bibnamefont{and} \bibinfo{author}{\bibfnamefont{G.}~\bibnamefont{Hiller}},
  \bibinfo{journal}{JHEP} \textbf{\bibinfo{volume}{06}}, \bibinfo{pages}{072}
  (\bibinfo{year}{2015}), \eprint{1503.01084}.

\bibitem[{\citenamefont{Crivellin
  et~al.}(2020{\natexlab{a}})\citenamefont{Crivellin, M\"uller, and
  Saturnino}}]{Crivellin:2019dwb}
\bibinfo{author}{\bibfnamefont{A.}~\bibnamefont{Crivellin}},
  \bibinfo{author}{\bibfnamefont{D.}~\bibnamefont{M\"uller}}, \bibnamefont{and}
  \bibinfo{author}{\bibfnamefont{F.}~\bibnamefont{Saturnino}},
  \bibinfo{journal}{JHEP} \textbf{\bibinfo{volume}{06}}, \bibinfo{pages}{020}
  (\bibinfo{year}{2020}{\natexlab{a}}), \eprint{1912.04224}.

\bibitem[{\citenamefont{Saad}(2020)}]{Saad:2020ihm}
\bibinfo{author}{\bibfnamefont{S.}~\bibnamefont{Saad}}, \bibinfo{journal}{Phys.
  Rev. D} \textbf{\bibinfo{volume}{102}}, \bibinfo{pages}{015019}
  (\bibinfo{year}{2020}), \eprint{2005.04352}.

\bibitem[{\citenamefont{Saad and Thapa}(2020)}]{Saad:2020ucl}
\bibinfo{author}{\bibfnamefont{S.}~\bibnamefont{Saad}} \bibnamefont{and}
  \bibinfo{author}{\bibfnamefont{A.}~\bibnamefont{Thapa}},
  \bibinfo{journal}{Phys. Rev. D} \textbf{\bibinfo{volume}{102}},
  \bibinfo{pages}{015014} (\bibinfo{year}{2020}), \eprint{2004.07880}.

\bibitem[{\citenamefont{Gherardi et~al.}(2021)\citenamefont{Gherardi, Marzocca,
  and Venturini}}]{Gherardi:2020qhc}
\bibinfo{author}{\bibfnamefont{V.}~\bibnamefont{Gherardi}},
  \bibinfo{author}{\bibfnamefont{D.}~\bibnamefont{Marzocca}}, \bibnamefont{and}
  \bibinfo{author}{\bibfnamefont{E.}~\bibnamefont{Venturini}},
  \bibinfo{journal}{JHEP} \textbf{\bibinfo{volume}{01}}, \bibinfo{pages}{138}
  (\bibinfo{year}{2021}), \eprint{2008.09548}.

\bibitem[{\citenamefont{Da~Rold and Lamagna}(2021)}]{DaRold:2020bib}
\bibinfo{author}{\bibfnamefont{L.}~\bibnamefont{Da~Rold}} \bibnamefont{and}
  \bibinfo{author}{\bibfnamefont{F.}~\bibnamefont{Lamagna}},
  \bibinfo{journal}{Phys. Rev. D} \textbf{\bibinfo{volume}{103}},
  \bibinfo{pages}{115007} (\bibinfo{year}{2021}), \eprint{2011.10061}.

\bibitem[{\citenamefont{Heeck and Thapa}(2022)}]{Heeck:2022znj}
\bibinfo{author}{\bibfnamefont{J.}~\bibnamefont{Heeck}} \bibnamefont{and}
  \bibinfo{author}{\bibfnamefont{A.}~\bibnamefont{Thapa}}
  (\bibinfo{year}{2022}), \eprint{2202.08854}.

\bibitem[{\citenamefont{Gripaios et~al.}(2016)\citenamefont{Gripaios,
  Nardecchia, and Renner}}]{Gripaios:2015gra}
\bibinfo{author}{\bibfnamefont{B.}~\bibnamefont{Gripaios}},
  \bibinfo{author}{\bibfnamefont{M.}~\bibnamefont{Nardecchia}},
  \bibnamefont{and} \bibinfo{author}{\bibfnamefont{S.~A.}
  \bibnamefont{Renner}}, \bibinfo{journal}{JHEP} \textbf{\bibinfo{volume}{06}},
  \bibinfo{pages}{083} (\bibinfo{year}{2016}), \eprint{1509.05020}.

\bibitem[{\citenamefont{Arnan et~al.}(2017)\citenamefont{Arnan, Hofer, Mescia,
  and Crivellin}}]{Arnan:2016cpy}
\bibinfo{author}{\bibfnamefont{P.}~\bibnamefont{Arnan}},
  \bibinfo{author}{\bibfnamefont{L.}~\bibnamefont{Hofer}},
  \bibinfo{author}{\bibfnamefont{F.}~\bibnamefont{Mescia}}, \bibnamefont{and}
  \bibinfo{author}{\bibfnamefont{A.}~\bibnamefont{Crivellin}},
  \bibinfo{journal}{JHEP} \textbf{\bibinfo{volume}{04}}, \bibinfo{pages}{043}
  (\bibinfo{year}{2017}), \eprint{1608.07832}.

\bibitem[{\citenamefont{Grinstein et~al.}(2018)\citenamefont{Grinstein,
  Pokorski, and Ross}}]{Grinstein:2018fgb}
\bibinfo{author}{\bibfnamefont{B.}~\bibnamefont{Grinstein}},
  \bibinfo{author}{\bibfnamefont{S.}~\bibnamefont{Pokorski}}, \bibnamefont{and}
  \bibinfo{author}{\bibfnamefont{G.~G.} \bibnamefont{Ross}},
  \bibinfo{journal}{JHEP} \textbf{\bibinfo{volume}{12}}, \bibinfo{pages}{079}
  (\bibinfo{year}{2018}), \eprint{1809.01766}.

\bibitem[{\citenamefont{Li et~al.}(2018)\citenamefont{Li, Li, Yang, and
  Zhang}}]{Li:2018rax}
\bibinfo{author}{\bibfnamefont{S.-P.} \bibnamefont{Li}},
  \bibinfo{author}{\bibfnamefont{X.-Q.} \bibnamefont{Li}},
  \bibinfo{author}{\bibfnamefont{Y.-D.} \bibnamefont{Yang}}, \bibnamefont{and}
  \bibinfo{author}{\bibfnamefont{X.}~\bibnamefont{Zhang}},
  \bibinfo{journal}{JHEP} \textbf{\bibinfo{volume}{09}}, \bibinfo{pages}{149}
  (\bibinfo{year}{2018}), \eprint{1807.08530}.

\bibitem[{\citenamefont{Marzo et~al.}(2019)\citenamefont{Marzo, Marzola, and
  Raidal}}]{Marzo:2019ldg}
\bibinfo{author}{\bibfnamefont{C.}~\bibnamefont{Marzo}},
  \bibinfo{author}{\bibfnamefont{L.}~\bibnamefont{Marzola}}, \bibnamefont{and}
  \bibinfo{author}{\bibfnamefont{M.}~\bibnamefont{Raidal}},
  \bibinfo{journal}{Phys. Rev. D} \textbf{\bibinfo{volume}{100}},
  \bibinfo{pages}{055031} (\bibinfo{year}{2019}), \eprint{1901.08290}.

\bibitem[{\citenamefont{Crivellin
  et~al.}(2019{\natexlab{b}})\citenamefont{Crivellin, M\"uller, and
  Wiegand}}]{Crivellin:2019dun}
\bibinfo{author}{\bibfnamefont{A.}~\bibnamefont{Crivellin}},
  \bibinfo{author}{\bibfnamefont{D.}~\bibnamefont{M\"uller}}, \bibnamefont{and}
  \bibinfo{author}{\bibfnamefont{C.}~\bibnamefont{Wiegand}},
  \bibinfo{journal}{JHEP} \textbf{\bibinfo{volume}{06}}, \bibinfo{pages}{119}
  (\bibinfo{year}{2019}{\natexlab{b}}), \eprint{1903.10440}.

\bibitem[{\citenamefont{Arnan et~al.}(2019)\citenamefont{Arnan, Crivellin,
  Fedele, and Mescia}}]{Arnan:2019uhr}
\bibinfo{author}{\bibfnamefont{P.}~\bibnamefont{Arnan}},
  \bibinfo{author}{\bibfnamefont{A.}~\bibnamefont{Crivellin}},
  \bibinfo{author}{\bibfnamefont{M.}~\bibnamefont{Fedele}}, \bibnamefont{and}
  \bibinfo{author}{\bibfnamefont{F.}~\bibnamefont{Mescia}},
  \bibinfo{journal}{JHEP} \textbf{\bibinfo{volume}{06}}, \bibinfo{pages}{118}
  (\bibinfo{year}{2019}), \eprint{1904.05890}.

\bibitem[{\citenamefont{Buras and Girrbach}(2013)}]{Buras:2013qja}
\bibinfo{author}{\bibfnamefont{A.~J.} \bibnamefont{Buras}} \bibnamefont{and}
  \bibinfo{author}{\bibfnamefont{J.}~\bibnamefont{Girrbach}},
  \bibinfo{journal}{JHEP} \textbf{\bibinfo{volume}{12}}, \bibinfo{pages}{009}
  (\bibinfo{year}{2013}), \eprint{1309.2466}.

\bibitem[{\citenamefont{Gauld et~al.}(2014{\natexlab{a}})\citenamefont{Gauld,
  Goertz, and Haisch}}]{Gauld:2013qba}
\bibinfo{author}{\bibfnamefont{R.}~\bibnamefont{Gauld}},
  \bibinfo{author}{\bibfnamefont{F.}~\bibnamefont{Goertz}}, \bibnamefont{and}
  \bibinfo{author}{\bibfnamefont{U.}~\bibnamefont{Haisch}},
  \bibinfo{journal}{Phys. Rev. D} \textbf{\bibinfo{volume}{89}},
  \bibinfo{pages}{015005} (\bibinfo{year}{2014}{\natexlab{a}}),
  \eprint{1308.1959}.

\bibitem[{\citenamefont{Gauld et~al.}(2014{\natexlab{b}})\citenamefont{Gauld,
  Goertz, and Haisch}}]{Gauld:2013qja}
\bibinfo{author}{\bibfnamefont{R.}~\bibnamefont{Gauld}},
  \bibinfo{author}{\bibfnamefont{F.}~\bibnamefont{Goertz}}, \bibnamefont{and}
  \bibinfo{author}{\bibfnamefont{U.}~\bibnamefont{Haisch}},
  \bibinfo{journal}{JHEP} \textbf{\bibinfo{volume}{01}}, \bibinfo{pages}{069}
  (\bibinfo{year}{2014}{\natexlab{b}}), \eprint{1310.1082}.

\bibitem[{\citenamefont{Altmannshofer et~al.}(2014)\citenamefont{Altmannshofer,
  Gori, Pospelov, and Yavin}}]{Altmannshofer:2014cfa}
\bibinfo{author}{\bibfnamefont{W.}~\bibnamefont{Altmannshofer}},
  \bibinfo{author}{\bibfnamefont{S.}~\bibnamefont{Gori}},
  \bibinfo{author}{\bibfnamefont{M.}~\bibnamefont{Pospelov}}, \bibnamefont{and}
  \bibinfo{author}{\bibfnamefont{I.}~\bibnamefont{Yavin}},
  \bibinfo{journal}{Phys. Rev. D} \textbf{\bibinfo{volume}{89}},
  \bibinfo{pages}{095033} (\bibinfo{year}{2014}), \eprint{1403.1269}.

\bibitem[{\citenamefont{Crivellin
  et~al.}(2015{\natexlab{a}})\citenamefont{Crivellin, D'Ambrosio, and
  Heeck}}]{Crivellin:2015mga}
\bibinfo{author}{\bibfnamefont{A.}~\bibnamefont{Crivellin}},
  \bibinfo{author}{\bibfnamefont{G.}~\bibnamefont{D'Ambrosio}},
  \bibnamefont{and} \bibinfo{author}{\bibfnamefont{J.}~\bibnamefont{Heeck}},
  \bibinfo{journal}{Phys. Rev. Lett.} \textbf{\bibinfo{volume}{114}},
  \bibinfo{pages}{151801} (\bibinfo{year}{2015}{\natexlab{a}}),
  \eprint{1501.00993}.

\bibitem[{\citenamefont{Crivellin
  et~al.}(2015{\natexlab{b}})\citenamefont{Crivellin, D'Ambrosio, and
  Heeck}}]{Crivellin:2015lwa}
\bibinfo{author}{\bibfnamefont{A.}~\bibnamefont{Crivellin}},
  \bibinfo{author}{\bibfnamefont{G.}~\bibnamefont{D'Ambrosio}},
  \bibnamefont{and} \bibinfo{author}{\bibfnamefont{J.}~\bibnamefont{Heeck}},
  \bibinfo{journal}{Phys. Rev. D} \textbf{\bibinfo{volume}{91}},
  \bibinfo{pages}{075006} (\bibinfo{year}{2015}{\natexlab{b}}),
  \eprint{1503.03477}.

\bibitem[{\citenamefont{Niehoff et~al.}(2015)\citenamefont{Niehoff, Stangl, and
  Straub}}]{Niehoff:2015bfa}
\bibinfo{author}{\bibfnamefont{C.}~\bibnamefont{Niehoff}},
  \bibinfo{author}{\bibfnamefont{P.}~\bibnamefont{Stangl}}, \bibnamefont{and}
  \bibinfo{author}{\bibfnamefont{D.~M.} \bibnamefont{Straub}},
  \bibinfo{journal}{Phys. Lett. B} \textbf{\bibinfo{volume}{747}},
  \bibinfo{pages}{182} (\bibinfo{year}{2015}), \eprint{1503.03865}.

\bibitem[{\citenamefont{Aristizabal~Sierra
  et~al.}(2015)\citenamefont{Aristizabal~Sierra, Staub, and
  Vicente}}]{Sierra:2015fma}
\bibinfo{author}{\bibfnamefont{D.}~\bibnamefont{Aristizabal~Sierra}},
  \bibinfo{author}{\bibfnamefont{F.}~\bibnamefont{Staub}}, \bibnamefont{and}
  \bibinfo{author}{\bibfnamefont{A.}~\bibnamefont{Vicente}},
  \bibinfo{journal}{Phys. Rev. D} \textbf{\bibinfo{volume}{92}},
  \bibinfo{pages}{015001} (\bibinfo{year}{2015}), \eprint{1503.06077}.

\bibitem[{\citenamefont{Carmona and Goertz}(2016)}]{Carmona:2015ena}
\bibinfo{author}{\bibfnamefont{A.}~\bibnamefont{Carmona}} \bibnamefont{and}
  \bibinfo{author}{\bibfnamefont{F.}~\bibnamefont{Goertz}},
  \bibinfo{journal}{Phys. Rev. Lett.} \textbf{\bibinfo{volume}{116}},
  \bibinfo{pages}{251801} (\bibinfo{year}{2016}), \eprint{1510.07658}.

\bibitem[{\citenamefont{Falkowski et~al.}(2015)\citenamefont{Falkowski,
  Nardecchia, and Ziegler}}]{Falkowski:2015zwa}
\bibinfo{author}{\bibfnamefont{A.}~\bibnamefont{Falkowski}},
  \bibinfo{author}{\bibfnamefont{M.}~\bibnamefont{Nardecchia}},
  \bibnamefont{and} \bibinfo{author}{\bibfnamefont{R.}~\bibnamefont{Ziegler}},
  \bibinfo{journal}{JHEP} \textbf{\bibinfo{volume}{11}}, \bibinfo{pages}{173}
  (\bibinfo{year}{2015}), \eprint{1509.01249}.

\bibitem[{\citenamefont{Celis et~al.}(2016)\citenamefont{Celis, Feng, and
  L\"ust}}]{Celis:2015eqs}
\bibinfo{author}{\bibfnamefont{A.}~\bibnamefont{Celis}},
  \bibinfo{author}{\bibfnamefont{W.-Z.} \bibnamefont{Feng}}, \bibnamefont{and}
  \bibinfo{author}{\bibfnamefont{D.}~\bibnamefont{L\"ust}},
  \bibinfo{journal}{JHEP} \textbf{\bibinfo{volume}{02}}, \bibinfo{pages}{007}
  (\bibinfo{year}{2016}), \eprint{1512.02218}.

\bibitem[{\citenamefont{Celis et~al.}(2015)\citenamefont{Celis, Fuentes-Martin,
  Jung, and Serodio}}]{Celis:2015ara}
\bibinfo{author}{\bibfnamefont{A.}~\bibnamefont{Celis}},
  \bibinfo{author}{\bibfnamefont{J.}~\bibnamefont{Fuentes-Martin}},
  \bibinfo{author}{\bibfnamefont{M.}~\bibnamefont{Jung}}, \bibnamefont{and}
  \bibinfo{author}{\bibfnamefont{H.}~\bibnamefont{Serodio}},
  \bibinfo{journal}{Phys. Rev. D} \textbf{\bibinfo{volume}{92}},
  \bibinfo{pages}{015007} (\bibinfo{year}{2015}), \eprint{1505.03079}.

\bibitem[{\citenamefont{Crivellin
  et~al.}(2015{\natexlab{c}})\citenamefont{Crivellin, Hofer, Matias, Nierste,
  Pokorski, and Rosiek}}]{Crivellin:2015era}
\bibinfo{author}{\bibfnamefont{A.}~\bibnamefont{Crivellin}},
  \bibinfo{author}{\bibfnamefont{L.}~\bibnamefont{Hofer}},
  \bibinfo{author}{\bibfnamefont{J.}~\bibnamefont{Matias}},
  \bibinfo{author}{\bibfnamefont{U.}~\bibnamefont{Nierste}},
  \bibinfo{author}{\bibfnamefont{S.}~\bibnamefont{Pokorski}}, \bibnamefont{and}
  \bibinfo{author}{\bibfnamefont{J.}~\bibnamefont{Rosiek}},
  \bibinfo{journal}{Phys. Rev. D} \textbf{\bibinfo{volume}{92}},
  \bibinfo{pages}{054013} (\bibinfo{year}{2015}{\natexlab{c}}),
  \eprint{1504.07928}.

\bibitem[{\citenamefont{Boucenna
  et~al.}(2016{\natexlab{a}})\citenamefont{Boucenna, Celis, Fuentes-Martin,
  Vicente, and Virto}}]{Boucenna:2016wpr}
\bibinfo{author}{\bibfnamefont{S.~M.} \bibnamefont{Boucenna}},
  \bibinfo{author}{\bibfnamefont{A.}~\bibnamefont{Celis}},
  \bibinfo{author}{\bibfnamefont{J.}~\bibnamefont{Fuentes-Martin}},
  \bibinfo{author}{\bibfnamefont{A.}~\bibnamefont{Vicente}}, \bibnamefont{and}
  \bibinfo{author}{\bibfnamefont{J.}~\bibnamefont{Virto}},
  \bibinfo{journal}{Phys. Lett. B} \textbf{\bibinfo{volume}{760}},
  \bibinfo{pages}{214} (\bibinfo{year}{2016}{\natexlab{a}}),
  \eprint{1604.03088}.

\bibitem[{\citenamefont{Altmannshofer et~al.}(2016)\citenamefont{Altmannshofer,
  Carena, and Crivellin}}]{Altmannshofer:2016oaq}
\bibinfo{author}{\bibfnamefont{W.}~\bibnamefont{Altmannshofer}},
  \bibinfo{author}{\bibfnamefont{M.}~\bibnamefont{Carena}}, \bibnamefont{and}
  \bibinfo{author}{\bibfnamefont{A.}~\bibnamefont{Crivellin}},
  \bibinfo{journal}{Phys. Rev. D} \textbf{\bibinfo{volume}{94}},
  \bibinfo{pages}{095026} (\bibinfo{year}{2016}), \eprint{1604.08221}.

\bibitem[{\citenamefont{Boucenna
  et~al.}(2016{\natexlab{b}})\citenamefont{Boucenna, Celis, Fuentes-Martin,
  Vicente, and Virto}}]{Boucenna:2016qad}
\bibinfo{author}{\bibfnamefont{S.~M.} \bibnamefont{Boucenna}},
  \bibinfo{author}{\bibfnamefont{A.}~\bibnamefont{Celis}},
  \bibinfo{author}{\bibfnamefont{J.}~\bibnamefont{Fuentes-Martin}},
  \bibinfo{author}{\bibfnamefont{A.}~\bibnamefont{Vicente}}, \bibnamefont{and}
  \bibinfo{author}{\bibfnamefont{J.}~\bibnamefont{Virto}},
  \bibinfo{journal}{JHEP} \textbf{\bibinfo{volume}{12}}, \bibinfo{pages}{059}
  (\bibinfo{year}{2016}{\natexlab{b}}), \eprint{1608.01349}.

\bibitem[{\citenamefont{Crivellin et~al.}(2017)\citenamefont{Crivellin,
  Fuentes-Martin, Greljo, and Isidori}}]{Crivellin:2016ejn}
\bibinfo{author}{\bibfnamefont{A.}~\bibnamefont{Crivellin}},
  \bibinfo{author}{\bibfnamefont{J.}~\bibnamefont{Fuentes-Martin}},
  \bibinfo{author}{\bibfnamefont{A.}~\bibnamefont{Greljo}}, \bibnamefont{and}
  \bibinfo{author}{\bibfnamefont{G.}~\bibnamefont{Isidori}},
  \bibinfo{journal}{Phys. Lett. B} \textbf{\bibinfo{volume}{766}},
  \bibinfo{pages}{77} (\bibinfo{year}{2017}), \eprint{1611.02703}.

\bibitem[{\citenamefont{Garcia~Garcia}(2017)}]{GarciaGarcia:2016nvr}
\bibinfo{author}{\bibfnamefont{I.}~\bibnamefont{Garcia~Garcia}},
  \bibinfo{journal}{JHEP} \textbf{\bibinfo{volume}{03}}, \bibinfo{pages}{040}
  (\bibinfo{year}{2017}), \eprint{1611.03507}.

\bibitem[{\citenamefont{Faisel and Tandean}(2018)}]{Faisel:2017glo}
\bibinfo{author}{\bibfnamefont{G.}~\bibnamefont{Faisel}} \bibnamefont{and}
  \bibinfo{author}{\bibfnamefont{J.}~\bibnamefont{Tandean}},
  \bibinfo{journal}{JHEP} \textbf{\bibinfo{volume}{02}}, \bibinfo{pages}{074}
  (\bibinfo{year}{2018}), \eprint{1710.11102}.

\bibitem[{\citenamefont{King}(2017)}]{King:2017anf}
\bibinfo{author}{\bibfnamefont{S.~F.} \bibnamefont{King}},
  \bibinfo{journal}{JHEP} \textbf{\bibinfo{volume}{08}}, \bibinfo{pages}{019}
  (\bibinfo{year}{2017}), \eprint{1706.06100}.

\bibitem[{\citenamefont{Chiang et~al.}(2017)\citenamefont{Chiang, He, Tandean,
  and Yuan}}]{Chiang:2017hlj}
\bibinfo{author}{\bibfnamefont{C.-W.} \bibnamefont{Chiang}},
  \bibinfo{author}{\bibfnamefont{X.-G.} \bibnamefont{He}},
  \bibinfo{author}{\bibfnamefont{J.}~\bibnamefont{Tandean}}, \bibnamefont{and}
  \bibinfo{author}{\bibfnamefont{X.-B.} \bibnamefont{Yuan}},
  \bibinfo{journal}{Phys. Rev. D} \textbf{\bibinfo{volume}{96}},
  \bibinfo{pages}{115022} (\bibinfo{year}{2017}), \eprint{1706.02696}.

\bibitem[{\citenamefont{Di~Chiara et~al.}(2017)\citenamefont{Di~Chiara, Fowlie,
  Fraser, Marzo, Marzola, Raidal, and Spethmann}}]{DiChiara:2017cjq}
\bibinfo{author}{\bibfnamefont{S.}~\bibnamefont{Di~Chiara}},
  \bibinfo{author}{\bibfnamefont{A.}~\bibnamefont{Fowlie}},
  \bibinfo{author}{\bibfnamefont{S.}~\bibnamefont{Fraser}},
  \bibinfo{author}{\bibfnamefont{C.}~\bibnamefont{Marzo}},
  \bibinfo{author}{\bibfnamefont{L.}~\bibnamefont{Marzola}},
  \bibinfo{author}{\bibfnamefont{M.}~\bibnamefont{Raidal}}, \bibnamefont{and}
  \bibinfo{author}{\bibfnamefont{C.}~\bibnamefont{Spethmann}},
  \bibinfo{journal}{Nucl. Phys. B} \textbf{\bibinfo{volume}{923}},
  \bibinfo{pages}{245} (\bibinfo{year}{2017}), \eprint{1704.06200}.

\bibitem[{\citenamefont{Ko et~al.}(2017)\citenamefont{Ko, Omura, Shigekami, and
  Yu}}]{Ko:2017lzd}
\bibinfo{author}{\bibfnamefont{P.}~\bibnamefont{Ko}},
  \bibinfo{author}{\bibfnamefont{Y.}~\bibnamefont{Omura}},
  \bibinfo{author}{\bibfnamefont{Y.}~\bibnamefont{Shigekami}},
  \bibnamefont{and} \bibinfo{author}{\bibfnamefont{C.}~\bibnamefont{Yu}},
  \bibinfo{journal}{Phys. Rev. D} \textbf{\bibinfo{volume}{95}},
  \bibinfo{pages}{115040} (\bibinfo{year}{2017}), \eprint{1702.08666}.

\bibitem[{\citenamefont{Sannino et~al.}(2018)\citenamefont{Sannino, Stangl,
  Straub, and Thomsen}}]{Sannino:2017utc}
\bibinfo{author}{\bibfnamefont{F.}~\bibnamefont{Sannino}},
  \bibinfo{author}{\bibfnamefont{P.}~\bibnamefont{Stangl}},
  \bibinfo{author}{\bibfnamefont{D.~M.} \bibnamefont{Straub}},
  \bibnamefont{and} \bibinfo{author}{\bibfnamefont{A.~E.}
  \bibnamefont{Thomsen}}, \bibinfo{journal}{Phys. Rev. D}
  \textbf{\bibinfo{volume}{97}}, \bibinfo{pages}{115046}
  (\bibinfo{year}{2018}), \eprint{1712.07646}.

\bibitem[{\citenamefont{Raby and Trautner}(2018)}]{Raby:2017igl}
\bibinfo{author}{\bibfnamefont{S.}~\bibnamefont{Raby}} \bibnamefont{and}
  \bibinfo{author}{\bibfnamefont{A.}~\bibnamefont{Trautner}},
  \bibinfo{journal}{Phys. Rev. D} \textbf{\bibinfo{volume}{97}},
  \bibinfo{pages}{095006} (\bibinfo{year}{2018}), \eprint{1712.09360}.

\bibitem[{\citenamefont{Alonso et~al.}(2017)\citenamefont{Alonso, Cox, Han, and
  Yanagida}}]{Alonso:2017bff}
\bibinfo{author}{\bibfnamefont{R.}~\bibnamefont{Alonso}},
  \bibinfo{author}{\bibfnamefont{P.}~\bibnamefont{Cox}},
  \bibinfo{author}{\bibfnamefont{C.}~\bibnamefont{Han}}, \bibnamefont{and}
  \bibinfo{author}{\bibfnamefont{T.~T.} \bibnamefont{Yanagida}},
  \bibinfo{journal}{Phys. Rev. D} \textbf{\bibinfo{volume}{96}},
  \bibinfo{pages}{071701} (\bibinfo{year}{2017}), \eprint{1704.08158}.

\bibitem[{\citenamefont{Cline and Martin~Camalich}(2017)}]{Cline:2017ihf}
\bibinfo{author}{\bibfnamefont{J.~M.} \bibnamefont{Cline}} \bibnamefont{and}
  \bibinfo{author}{\bibfnamefont{J.}~\bibnamefont{Martin~Camalich}},
  \bibinfo{journal}{Phys. Rev. D} \textbf{\bibinfo{volume}{96}},
  \bibinfo{pages}{055036} (\bibinfo{year}{2017}), \eprint{1706.08510}.

\bibitem[{\citenamefont{Carmona and Goertz}(2018)}]{Carmona:2017fsn}
\bibinfo{author}{\bibfnamefont{A.}~\bibnamefont{Carmona}} \bibnamefont{and}
  \bibinfo{author}{\bibfnamefont{F.}~\bibnamefont{Goertz}},
  \bibinfo{journal}{Eur. Phys. J. C} \textbf{\bibinfo{volume}{78}},
  \bibinfo{pages}{979} (\bibinfo{year}{2018}), \eprint{1712.02536}.

\bibitem[{\citenamefont{Falkowski et~al.}(2018)\citenamefont{Falkowski, King,
  Perdomo, and Pierre}}]{Falkowski:2018dsl}
\bibinfo{author}{\bibfnamefont{A.}~\bibnamefont{Falkowski}},
  \bibinfo{author}{\bibfnamefont{S.~F.} \bibnamefont{King}},
  \bibinfo{author}{\bibfnamefont{E.}~\bibnamefont{Perdomo}}, \bibnamefont{and}
  \bibinfo{author}{\bibfnamefont{M.}~\bibnamefont{Pierre}},
  \bibinfo{journal}{JHEP} \textbf{\bibinfo{volume}{08}}, \bibinfo{pages}{061}
  (\bibinfo{year}{2018}), \eprint{1803.04430}.

\bibitem[{\citenamefont{Benavides et~al.}(2020)\citenamefont{Benavides,
  Mu\~noz, Ponce, Rodr\'\i{}guez, and Rojas}}]{Benavides:2018rgh}
\bibinfo{author}{\bibfnamefont{R.~H.} \bibnamefont{Benavides}},
  \bibinfo{author}{\bibfnamefont{L.}~\bibnamefont{Mu\~noz}},
  \bibinfo{author}{\bibfnamefont{W.~A.} \bibnamefont{Ponce}},
  \bibinfo{author}{\bibfnamefont{O.}~\bibnamefont{Rodr\'\i{}guez}},
  \bibnamefont{and} \bibinfo{author}{\bibfnamefont{E.}~\bibnamefont{Rojas}},
  \bibinfo{journal}{J. Phys. G} \textbf{\bibinfo{volume}{47}},
  \bibinfo{pages}{075003} (\bibinfo{year}{2020}), \eprint{1812.05077}.

\bibitem[{\citenamefont{Maji et~al.}(2019)\citenamefont{Maji, Nayek, and
  Sahoo}}]{Maji:2018gvz}
\bibinfo{author}{\bibfnamefont{P.}~\bibnamefont{Maji}},
  \bibinfo{author}{\bibfnamefont{P.}~\bibnamefont{Nayek}}, \bibnamefont{and}
  \bibinfo{author}{\bibfnamefont{S.}~\bibnamefont{Sahoo}},
  \bibinfo{journal}{PTEP} \textbf{\bibinfo{volume}{2019}},
  \bibinfo{pages}{033B06} (\bibinfo{year}{2019}), \eprint{1811.03869}.

\bibitem[{\citenamefont{Singirala et~al.}(2019)\citenamefont{Singirala, Sahoo,
  and Mohanta}}]{Singirala:2018mio}
\bibinfo{author}{\bibfnamefont{S.}~\bibnamefont{Singirala}},
  \bibinfo{author}{\bibfnamefont{S.}~\bibnamefont{Sahoo}}, \bibnamefont{and}
  \bibinfo{author}{\bibfnamefont{R.}~\bibnamefont{Mohanta}},
  \bibinfo{journal}{Phys. Rev. D} \textbf{\bibinfo{volume}{99}},
  \bibinfo{pages}{035042} (\bibinfo{year}{2019}), \eprint{1809.03213}.

\bibitem[{\citenamefont{Guadagnoli et~al.}(2018)\citenamefont{Guadagnoli,
  Reboud, and Sumensari}}]{Guadagnoli:2018ojc}
\bibinfo{author}{\bibfnamefont{D.}~\bibnamefont{Guadagnoli}},
  \bibinfo{author}{\bibfnamefont{M.}~\bibnamefont{Reboud}}, \bibnamefont{and}
  \bibinfo{author}{\bibfnamefont{O.}~\bibnamefont{Sumensari}},
  \bibinfo{journal}{JHEP} \textbf{\bibinfo{volume}{11}}, \bibinfo{pages}{163}
  (\bibinfo{year}{2018}), \eprint{1807.03285}.

\bibitem[{\citenamefont{Allanach and Davighi}(2018)}]{Allanach:2018lvl}
\bibinfo{author}{\bibfnamefont{B.~C.} \bibnamefont{Allanach}} \bibnamefont{and}
  \bibinfo{author}{\bibfnamefont{J.}~\bibnamefont{Davighi}},
  \bibinfo{journal}{JHEP} \textbf{\bibinfo{volume}{12}}, \bibinfo{pages}{075}
  (\bibinfo{year}{2018}), \eprint{1809.01158}.

\bibitem[{\citenamefont{Kohda et~al.}(2018)\citenamefont{Kohda, Modak, and
  Soffer}}]{Kohda:2018xbc}
\bibinfo{author}{\bibfnamefont{M.}~\bibnamefont{Kohda}},
  \bibinfo{author}{\bibfnamefont{T.}~\bibnamefont{Modak}}, \bibnamefont{and}
  \bibinfo{author}{\bibfnamefont{A.}~\bibnamefont{Soffer}},
  \bibinfo{journal}{Phys. Rev. D} \textbf{\bibinfo{volume}{97}},
  \bibinfo{pages}{115019} (\bibinfo{year}{2018}), \eprint{1803.07492}.

\bibitem[{\citenamefont{King}(2018)}]{King:2018fcg}
\bibinfo{author}{\bibfnamefont{S.~F.} \bibnamefont{King}},
  \bibinfo{journal}{JHEP} \textbf{\bibinfo{volume}{09}}, \bibinfo{pages}{069}
  (\bibinfo{year}{2018}), \eprint{1806.06780}.

\bibitem[{\citenamefont{Duan et~al.}(2019)\citenamefont{Duan, Fan, Frank, Han,
  and Yang}}]{Duan:2018akc}
\bibinfo{author}{\bibfnamefont{G.~H.} \bibnamefont{Duan}},
  \bibinfo{author}{\bibfnamefont{X.}~\bibnamefont{Fan}},
  \bibinfo{author}{\bibfnamefont{M.}~\bibnamefont{Frank}},
  \bibinfo{author}{\bibfnamefont{C.}~\bibnamefont{Han}}, \bibnamefont{and}
  \bibinfo{author}{\bibfnamefont{J.~M.} \bibnamefont{Yang}},
  \bibinfo{journal}{Phys. Lett. B} \textbf{\bibinfo{volume}{789}},
  \bibinfo{pages}{54} (\bibinfo{year}{2019}), \eprint{1808.04116}.

\bibitem[{\citenamefont{Rocha-Moran and Vicente}(2019)}]{Rocha-Moran:2018jzu}
\bibinfo{author}{\bibfnamefont{P.}~\bibnamefont{Rocha-Moran}} \bibnamefont{and}
  \bibinfo{author}{\bibfnamefont{A.}~\bibnamefont{Vicente}},
  \bibinfo{journal}{Phys. Rev. D} \textbf{\bibinfo{volume}{99}},
  \bibinfo{pages}{035016} (\bibinfo{year}{2019}), \eprint{1810.02135}.

\bibitem[{\citenamefont{Dwivedi et~al.}(2020)\citenamefont{Dwivedi,
  Kumar~Ghosh, Falkowski, and Ghosh}}]{Dwivedi:2019uqd}
\bibinfo{author}{\bibfnamefont{S.}~\bibnamefont{Dwivedi}},
  \bibinfo{author}{\bibfnamefont{D.}~\bibnamefont{Kumar~Ghosh}},
  \bibinfo{author}{\bibfnamefont{A.}~\bibnamefont{Falkowski}},
  \bibnamefont{and} \bibinfo{author}{\bibfnamefont{N.}~\bibnamefont{Ghosh}},
  \bibinfo{journal}{Eur. Phys. J. C} \textbf{\bibinfo{volume}{80}},
  \bibinfo{pages}{263} (\bibinfo{year}{2020}), \eprint{1908.03031}.

\bibitem[{\citenamefont{Foldenauer}(2019)}]{Foldenauer:2019vgn}
\bibinfo{author}{\bibfnamefont{P.}~\bibnamefont{Foldenauer}}, Ph.D. thesis,
  \bibinfo{school}{U. Heidelberg (main)} (\bibinfo{year}{2019}).

\bibitem[{\citenamefont{Ko et~al.}(2019)\citenamefont{Ko, Nomura, and
  Yu}}]{Ko:2019tts}
\bibinfo{author}{\bibfnamefont{P.}~\bibnamefont{Ko}},
  \bibinfo{author}{\bibfnamefont{T.}~\bibnamefont{Nomura}}, \bibnamefont{and}
  \bibinfo{author}{\bibfnamefont{C.}~\bibnamefont{Yu}}, \bibinfo{journal}{JHEP}
  \textbf{\bibinfo{volume}{04}}, \bibinfo{pages}{102} (\bibinfo{year}{2019}),
  \eprint{1902.06107}.

\bibitem[{\citenamefont{Allanach and Davighi}(2019)}]{Allanach:2019iiy}
\bibinfo{author}{\bibfnamefont{B.~C.} \bibnamefont{Allanach}} \bibnamefont{and}
  \bibinfo{author}{\bibfnamefont{J.}~\bibnamefont{Davighi}},
  \bibinfo{journal}{Eur. Phys. J. C} \textbf{\bibinfo{volume}{79}},
  \bibinfo{pages}{908} (\bibinfo{year}{2019}), \eprint{1905.10327}.

\bibitem[{\citenamefont{Kawamura et~al.}(2019)\citenamefont{Kawamura, Raby, and
  Trautner}}]{Kawamura:2019rth}
\bibinfo{author}{\bibfnamefont{J.}~\bibnamefont{Kawamura}},
  \bibinfo{author}{\bibfnamefont{S.}~\bibnamefont{Raby}}, \bibnamefont{and}
  \bibinfo{author}{\bibfnamefont{A.}~\bibnamefont{Trautner}},
  \bibinfo{journal}{Phys. Rev. D} \textbf{\bibinfo{volume}{100}},
  \bibinfo{pages}{055030} (\bibinfo{year}{2019}), \eprint{1906.11297}.

\bibitem[{\citenamefont{Altmannshofer et~al.}(2020)\citenamefont{Altmannshofer,
  Davighi, and Nardecchia}}]{Altmannshofer:2019xda}
\bibinfo{author}{\bibfnamefont{W.}~\bibnamefont{Altmannshofer}},
  \bibinfo{author}{\bibfnamefont{J.}~\bibnamefont{Davighi}}, \bibnamefont{and}
  \bibinfo{author}{\bibfnamefont{M.}~\bibnamefont{Nardecchia}},
  \bibinfo{journal}{Phys. Rev. D} \textbf{\bibinfo{volume}{101}},
  \bibinfo{pages}{015004} (\bibinfo{year}{2020}), \eprint{1909.02021}.

\bibitem[{\citenamefont{Calibbi et~al.}(2020)\citenamefont{Calibbi, Crivellin,
  Kirk, Manzari, and Vernazza}}]{Calibbi:2019lvs}
\bibinfo{author}{\bibfnamefont{L.}~\bibnamefont{Calibbi}},
  \bibinfo{author}{\bibfnamefont{A.}~\bibnamefont{Crivellin}},
  \bibinfo{author}{\bibfnamefont{F.}~\bibnamefont{Kirk}},
  \bibinfo{author}{\bibfnamefont{C.~A.} \bibnamefont{Manzari}},
  \bibnamefont{and} \bibinfo{author}{\bibfnamefont{L.}~\bibnamefont{Vernazza}},
  \bibinfo{journal}{Phys. Rev. D} \textbf{\bibinfo{volume}{101}},
  \bibinfo{pages}{095003} (\bibinfo{year}{2020}), \eprint{1910.00014}.

\bibitem[{\citenamefont{Aebischer et~al.}(2020)\citenamefont{Aebischer, Buras,
  Cerd\`a-Sevilla, and De~Fazio}}]{Aebischer:2019blw}
\bibinfo{author}{\bibfnamefont{J.}~\bibnamefont{Aebischer}},
  \bibinfo{author}{\bibfnamefont{A.~J.} \bibnamefont{Buras}},
  \bibinfo{author}{\bibfnamefont{M.}~\bibnamefont{Cerd\`a-Sevilla}},
  \bibnamefont{and} \bibinfo{author}{\bibfnamefont{F.}~\bibnamefont{De~Fazio}},
  \bibinfo{journal}{JHEP} \textbf{\bibinfo{volume}{02}}, \bibinfo{pages}{183}
  (\bibinfo{year}{2020}), \eprint{1912.09308}.

\bibitem[{\citenamefont{Kawamura et~al.}(2020)\citenamefont{Kawamura, Raby, and
  Trautner}}]{Kawamura:2019hxp}
\bibinfo{author}{\bibfnamefont{J.}~\bibnamefont{Kawamura}},
  \bibinfo{author}{\bibfnamefont{S.}~\bibnamefont{Raby}}, \bibnamefont{and}
  \bibinfo{author}{\bibfnamefont{A.}~\bibnamefont{Trautner}},
  \bibinfo{journal}{Phys. Rev. D} \textbf{\bibinfo{volume}{101}},
  \bibinfo{pages}{035026} (\bibinfo{year}{2020}), \eprint{1911.11075}.

\bibitem[{\citenamefont{Crivellin et~al.}(2021)\citenamefont{Crivellin,
  Manzari, Alguero, and Matias}}]{Crivellin:2020oup}
\bibinfo{author}{\bibfnamefont{A.}~\bibnamefont{Crivellin}},
  \bibinfo{author}{\bibfnamefont{C.~A.} \bibnamefont{Manzari}},
  \bibinfo{author}{\bibfnamefont{M.}~\bibnamefont{Alguero}}, \bibnamefont{and}
  \bibinfo{author}{\bibfnamefont{J.}~\bibnamefont{Matias}},
  \bibinfo{journal}{Phys. Rev. Lett.} \textbf{\bibinfo{volume}{127}},
  \bibinfo{pages}{011801} (\bibinfo{year}{2021}), \eprint{2010.14504}.

\bibitem[{\citenamefont{Allanach}(2021)}]{Allanach:2020kss}
\bibinfo{author}{\bibfnamefont{B.~C.} \bibnamefont{Allanach}},
  \bibinfo{journal}{Eur. Phys. J. C} \textbf{\bibinfo{volume}{81}},
  \bibinfo{pages}{56} (\bibinfo{year}{2021}), \bibinfo{note}{[Erratum:
  Eur.Phys.J.C 81, 321 (2021)]}, \eprint{2009.02197}.

\bibitem[{\citenamefont{Capdevila et~al.}(2021)\citenamefont{Capdevila,
  Crivellin, Manzari, and Montull}}]{Capdevila:2020rrl}
\bibinfo{author}{\bibfnamefont{B.}~\bibnamefont{Capdevila}},
  \bibinfo{author}{\bibfnamefont{A.}~\bibnamefont{Crivellin}},
  \bibinfo{author}{\bibfnamefont{C.~A.} \bibnamefont{Manzari}},
  \bibnamefont{and} \bibinfo{author}{\bibfnamefont{M.}~\bibnamefont{Montull}},
  \bibinfo{journal}{Phys. Rev. D} \textbf{\bibinfo{volume}{103}},
  \bibinfo{pages}{015032} (\bibinfo{year}{2021}), \eprint{2005.13542}.

\bibitem[{\citenamefont{Greljo et~al.}(2021)\citenamefont{Greljo, Stangl, and
  Thomsen}}]{Greljo:2021xmg}
\bibinfo{author}{\bibfnamefont{A.}~\bibnamefont{Greljo}},
  \bibinfo{author}{\bibfnamefont{P.}~\bibnamefont{Stangl}}, \bibnamefont{and}
  \bibinfo{author}{\bibfnamefont{A.~E.} \bibnamefont{Thomsen}},
  \bibinfo{journal}{Phys. Lett. B} \textbf{\bibinfo{volume}{820}},
  \bibinfo{pages}{136554} (\bibinfo{year}{2021}), \eprint{2103.13991}.

\bibitem[{\citenamefont{Davighi}(2021)}]{Davighi:2021oel}
\bibinfo{author}{\bibfnamefont{J.}~\bibnamefont{Davighi}},
  \bibinfo{journal}{JHEP} \textbf{\bibinfo{volume}{08}}, \bibinfo{pages}{101}
  (\bibinfo{year}{2021}), \eprint{2105.06918}.

\bibitem[{\citenamefont{Allanach
  et~al.}(2021{\natexlab{a}})\citenamefont{Allanach, Camargo-Molina, and
  Davighi}}]{Allanach:2021kzj}
\bibinfo{author}{\bibfnamefont{B.~C.} \bibnamefont{Allanach}},
  \bibinfo{author}{\bibfnamefont{J.~E.} \bibnamefont{Camargo-Molina}},
  \bibnamefont{and} \bibinfo{author}{\bibfnamefont{J.}~\bibnamefont{Davighi}},
  \bibinfo{journal}{Eur. Phys. J. C} \textbf{\bibinfo{volume}{81}},
  \bibinfo{pages}{721} (\bibinfo{year}{2021}{\natexlab{a}}),
  \eprint{2103.12056}.

\bibitem[{\citenamefont{Navarro and King}(2022)}]{Navarro:2021sfb}
\bibinfo{author}{\bibfnamefont{M.~F.} \bibnamefont{Navarro}} \bibnamefont{and}
  \bibinfo{author}{\bibfnamefont{S.~F.} \bibnamefont{King}},
  \bibinfo{journal}{Phys. Rev. D} \textbf{\bibinfo{volume}{105}},
  \bibinfo{pages}{035015} (\bibinfo{year}{2022}), \eprint{2109.08729}.

\bibitem[{\citenamefont{Ko et~al.}(2021)\citenamefont{Ko, Nomura, and
  Okada}}]{Ko:2021lpx}
\bibinfo{author}{\bibfnamefont{P.}~\bibnamefont{Ko}},
  \bibinfo{author}{\bibfnamefont{T.}~\bibnamefont{Nomura}}, \bibnamefont{and}
  \bibinfo{author}{\bibfnamefont{H.}~\bibnamefont{Okada}}
  (\bibinfo{year}{2021}), \eprint{2110.10513}.

\bibitem[{\citenamefont{Bause et~al.}(2022)\citenamefont{Bause, Hiller,
  H\"ohne, Litim, and Steudtner}}]{Bause:2021prv}
\bibinfo{author}{\bibfnamefont{R.}~\bibnamefont{Bause}},
  \bibinfo{author}{\bibfnamefont{G.}~\bibnamefont{Hiller}},
  \bibinfo{author}{\bibfnamefont{T.}~\bibnamefont{H\"ohne}},
  \bibinfo{author}{\bibfnamefont{D.~F.} \bibnamefont{Litim}}, \bibnamefont{and}
  \bibinfo{author}{\bibfnamefont{T.}~\bibnamefont{Steudtner}},
  \bibinfo{journal}{Eur. Phys. J. C} \textbf{\bibinfo{volume}{82}},
  \bibinfo{pages}{42} (\bibinfo{year}{2022}), \eprint{2109.06201}.

\bibitem[{\citenamefont{Allanach
  et~al.}(2021{\natexlab{b}})\citenamefont{Allanach, Butterworth, and
  Corbett}}]{Allanach:2021gmj}
\bibinfo{author}{\bibfnamefont{B.~C.} \bibnamefont{Allanach}},
  \bibinfo{author}{\bibfnamefont{J.~M.} \bibnamefont{Butterworth}},
  \bibnamefont{and} \bibinfo{author}{\bibfnamefont{T.}~\bibnamefont{Corbett}},
  \bibinfo{journal}{Eur. Phys. J. C} \textbf{\bibinfo{volume}{81}},
  \bibinfo{pages}{1126} (\bibinfo{year}{2021}{\natexlab{b}}),
  \eprint{2110.13518}.

\bibitem[{\citenamefont{Alguer\'o et~al.}(2022)\citenamefont{Alguer\'o,
  Crivellin, Manzari, and Matias}}]{Alguero:2022est}
\bibinfo{author}{\bibfnamefont{M.}~\bibnamefont{Alguer\'o}},
  \bibinfo{author}{\bibfnamefont{A.}~\bibnamefont{Crivellin}},
  \bibinfo{author}{\bibfnamefont{C.~A.} \bibnamefont{Manzari}},
  \bibnamefont{and} \bibinfo{author}{\bibfnamefont{J.}~\bibnamefont{Matias}}
  (\bibinfo{year}{2022}), \eprint{2201.08170}.

\bibitem[{\citenamefont{Sala and Straub}(2017)}]{Sala:2017ihs}
\bibinfo{author}{\bibfnamefont{F.}~\bibnamefont{Sala}} \bibnamefont{and}
  \bibinfo{author}{\bibfnamefont{D.~M.} \bibnamefont{Straub}},
  \bibinfo{journal}{Phys. Lett. B} \textbf{\bibinfo{volume}{774}},
  \bibinfo{pages}{205} (\bibinfo{year}{2017}), \eprint{1704.06188}.

\bibitem[{\citenamefont{Mohapatra and Giri}(2021)}]{Mohapatra:2021izl}
\bibinfo{author}{\bibfnamefont{M.~K.} \bibnamefont{Mohapatra}}
  \bibnamefont{and} \bibinfo{author}{\bibfnamefont{A.}~\bibnamefont{Giri}},
  \bibinfo{journal}{Phys. Rev. D} \textbf{\bibinfo{volume}{104}},
  \bibinfo{pages}{095012} (\bibinfo{year}{2021}), \eprint{2109.12382}.

\bibitem[{\citenamefont{Datta et~al.}(2018)\citenamefont{Datta, Kumar, Liao,
  and Marfatia}}]{Datta:2017ezo}
\bibinfo{author}{\bibfnamefont{A.}~\bibnamefont{Datta}},
  \bibinfo{author}{\bibfnamefont{J.}~\bibnamefont{Kumar}},
  \bibinfo{author}{\bibfnamefont{J.}~\bibnamefont{Liao}}, \bibnamefont{and}
  \bibinfo{author}{\bibfnamefont{D.}~\bibnamefont{Marfatia}},
  \bibinfo{journal}{Phys. Rev. D} \textbf{\bibinfo{volume}{97}},
  \bibinfo{pages}{115038} (\bibinfo{year}{2018}), \eprint{1705.08423}.

\bibitem[{\citenamefont{Altmannshofer et~al.}(2018)\citenamefont{Altmannshofer,
  Baker, Gori, Harnik, Pospelov, Stamou, and Thamm}}]{Altmannshofer:2017bsz}
\bibinfo{author}{\bibfnamefont{W.}~\bibnamefont{Altmannshofer}},
  \bibinfo{author}{\bibfnamefont{M.~J.} \bibnamefont{Baker}},
  \bibinfo{author}{\bibfnamefont{S.}~\bibnamefont{Gori}},
  \bibinfo{author}{\bibfnamefont{R.}~\bibnamefont{Harnik}},
  \bibinfo{author}{\bibfnamefont{M.}~\bibnamefont{Pospelov}},
  \bibinfo{author}{\bibfnamefont{E.}~\bibnamefont{Stamou}}, \bibnamefont{and}
  \bibinfo{author}{\bibfnamefont{A.}~\bibnamefont{Thamm}},
  \bibinfo{journal}{JHEP} \textbf{\bibinfo{volume}{03}}, \bibinfo{pages}{188}
  (\bibinfo{year}{2018}), \eprint{1711.07494}.

\bibitem[{\citenamefont{Sala}(2018)}]{Sala:2018ukk}
\bibinfo{author}{\bibfnamefont{F.}~\bibnamefont{Sala}}, \bibinfo{journal}{Nucl.
  Part. Phys. Proc.} \textbf{\bibinfo{volume}{303-305}}, \bibinfo{pages}{14}
  (\bibinfo{year}{2018}), \eprint{1809.11061}.

\bibitem[{\citenamefont{Bishara et~al.}(2017)\citenamefont{Bishara, Haisch, and
  Monni}}]{Bishara:2017pje}
\bibinfo{author}{\bibfnamefont{F.}~\bibnamefont{Bishara}},
  \bibinfo{author}{\bibfnamefont{U.}~\bibnamefont{Haisch}}, \bibnamefont{and}
  \bibinfo{author}{\bibfnamefont{P.~F.} \bibnamefont{Monni}},
  \bibinfo{journal}{Phys. Rev. D} \textbf{\bibinfo{volume}{96}},
  \bibinfo{pages}{055002} (\bibinfo{year}{2017}), \eprint{1705.03465}.

\bibitem[{\citenamefont{Borah et~al.}(2020)\citenamefont{Borah, Mukherjee, and
  Nandi}}]{Borah:2020swo}
\bibinfo{author}{\bibfnamefont{D.}~\bibnamefont{Borah}},
  \bibinfo{author}{\bibfnamefont{L.}~\bibnamefont{Mukherjee}},
  \bibnamefont{and} \bibinfo{author}{\bibfnamefont{S.}~\bibnamefont{Nandi}},
  \bibinfo{journal}{JHEP} \textbf{\bibinfo{volume}{12}}, \bibinfo{pages}{052}
  (\bibinfo{year}{2020}), \eprint{2007.13778}.

\bibitem[{\citenamefont{Darm\'e et~al.}(2021)\citenamefont{Darm\'e, Fedele,
  Kowalska, and Sessolo}}]{Darme:2021qzw}
\bibinfo{author}{\bibfnamefont{L.}~\bibnamefont{Darm\'e}},
  \bibinfo{author}{\bibfnamefont{M.}~\bibnamefont{Fedele}},
  \bibinfo{author}{\bibfnamefont{K.}~\bibnamefont{Kowalska}}, \bibnamefont{and}
  \bibinfo{author}{\bibfnamefont{E.~M.} \bibnamefont{Sessolo}}
  (\bibinfo{year}{2021}), \eprint{2106.12582}.

\bibitem[{\citenamefont{Adachi et~al.}(2020)}]{Adachi:2019otg}
\bibinfo{author}{\bibfnamefont{I.}~\bibnamefont{Adachi}} \bibnamefont{et~al.}
  (\bibinfo{collaboration}{Belle-II}), \bibinfo{journal}{Phys. Rev. Lett.}
  \textbf{\bibinfo{volume}{124}}, \bibinfo{pages}{141801}
  (\bibinfo{year}{2020}), \eprint{1912.11276}.

\bibitem[{\citenamefont{Grygier et~al.}(2017)}]{Belle:2017oht}
\bibinfo{author}{\bibfnamefont{J.}~\bibnamefont{Grygier}} \bibnamefont{et~al.}
  (\bibinfo{collaboration}{Belle}), \bibinfo{journal}{Phys. Rev. D}
  \textbf{\bibinfo{volume}{96}}, \bibinfo{pages}{091101}
  (\bibinfo{year}{2017}), \bibinfo{note}{[Addendum: Phys.Rev.D 97, 099902
  (2018)]}, \eprint{1702.03224}.

\bibitem[{\citenamefont{Straub}(2018)}]{Straub:2018kue}
\bibinfo{author}{\bibfnamefont{D.~M.} \bibnamefont{Straub}}
  (\bibinfo{year}{2018}), \eprint{1810.08132}.

\bibitem[{\citenamefont{Aaij et~al.}(2021{\natexlab{a}})}]{LHCb:2021trn}
\bibinfo{author}{\bibfnamefont{R.}~\bibnamefont{Aaij}} \bibnamefont{et~al.}
  (\bibinfo{collaboration}{LHCb}) (\bibinfo{year}{2021}{\natexlab{a}}),
  \eprint{2103.11769}.

\bibitem[{\citenamefont{Aaij et~al.}(2017)}]{LHCb:2017avl}
\bibinfo{author}{\bibfnamefont{R.}~\bibnamefont{Aaij}} \bibnamefont{et~al.}
  (\bibinfo{collaboration}{LHCb}), \bibinfo{journal}{JHEP}
  \textbf{\bibinfo{volume}{08}}, \bibinfo{pages}{055} (\bibinfo{year}{2017}),
  \eprint{1705.05802}.

\bibitem[{\citenamefont{Aaij et~al.}(2021{\natexlab{b}})}]{LHCb:2021lvy}
\bibinfo{author}{\bibfnamefont{R.}~\bibnamefont{Aaij}} \bibnamefont{et~al.}
  (\bibinfo{collaboration}{LHCb}) (\bibinfo{year}{2021}{\natexlab{b}}),
  \eprint{2110.09501}.

\bibitem[{\citenamefont{Khachatryan et~al.}(2015)}]{CMS:2014xfa}
\bibinfo{author}{\bibfnamefont{V.}~\bibnamefont{Khachatryan}}
  \bibnamefont{et~al.} (\bibinfo{collaboration}{CMS, LHCb}),
  \bibinfo{journal}{Nature} \textbf{\bibinfo{volume}{522}}, \bibinfo{pages}{68}
  (\bibinfo{year}{2015}), \eprint{1411.4413}.

\bibitem[{\citenamefont{Aaboud et~al.}(2019)}]{ATLAS:2018cur}
\bibinfo{author}{\bibfnamefont{M.}~\bibnamefont{Aaboud}} \bibnamefont{et~al.}
  (\bibinfo{collaboration}{ATLAS}), \bibinfo{journal}{JHEP}
  \textbf{\bibinfo{volume}{04}}, \bibinfo{pages}{098} (\bibinfo{year}{2019}),
  \eprint{1812.03017}.

\bibitem[{\citenamefont{Sirunyan et~al.}(2020)}]{CMS:2019bbr}
\bibinfo{author}{\bibfnamefont{A.~M.} \bibnamefont{Sirunyan}}
  \bibnamefont{et~al.} (\bibinfo{collaboration}{CMS}), \bibinfo{journal}{JHEP}
  \textbf{\bibinfo{volume}{04}}, \bibinfo{pages}{188} (\bibinfo{year}{2020}),
  \eprint{1910.12127}.

\bibitem[{\citenamefont{Aaij et~al.}(2021{\natexlab{c}})}]{LHCb:2020gog}
\bibinfo{author}{\bibfnamefont{R.}~\bibnamefont{Aaij}} \bibnamefont{et~al.}
  (\bibinfo{collaboration}{LHCb}), \bibinfo{journal}{Phys. Rev. Lett.}
  \textbf{\bibinfo{volume}{126}}, \bibinfo{pages}{161802}
  (\bibinfo{year}{2021}{\natexlab{c}}), \eprint{2012.13241}.

\bibitem[{\citenamefont{Aaij et~al.}(2021{\natexlab{d}})}]{LHCb:2021xxq}
\bibinfo{author}{\bibfnamefont{R.}~\bibnamefont{Aaij}} \bibnamefont{et~al.}
  (\bibinfo{collaboration}{LHCb}), \bibinfo{journal}{JHEP}
  \textbf{\bibinfo{volume}{11}}, \bibinfo{pages}{043}
  (\bibinfo{year}{2021}{\natexlab{d}}), \eprint{2107.13428}.

\bibitem[{\citenamefont{Aaij et~al.}(2021{\natexlab{e}})}]{LHCb:2021zwz}
\bibinfo{author}{\bibfnamefont{R.}~\bibnamefont{Aaij}} \bibnamefont{et~al.}
  (\bibinfo{collaboration}{LHCb}), \bibinfo{journal}{Phys. Rev. Lett.}
  \textbf{\bibinfo{volume}{127}}, \bibinfo{pages}{151801}
  (\bibinfo{year}{2021}{\natexlab{e}}), \eprint{2105.14007}.

\bibitem[{\citenamefont{Lees et~al.}(2013)}]{BaBar:2013npw}
\bibinfo{author}{\bibfnamefont{J.~P.} \bibnamefont{Lees}} \bibnamefont{et~al.}
  (\bibinfo{collaboration}{BaBar}), \bibinfo{journal}{Phys. Rev. D}
  \textbf{\bibinfo{volume}{87}}, \bibinfo{pages}{112005}
  (\bibinfo{year}{2013}), \eprint{1303.7465}.

\bibitem[{\citenamefont{Abudin\'en et~al.}(2021)}]{Belle-II:2021rof}
\bibinfo{author}{\bibfnamefont{F.}~\bibnamefont{Abudin\'en}}
  \bibnamefont{et~al.} (\bibinfo{collaboration}{Belle-II}),
  \bibinfo{journal}{Phys. Rev. Lett.} \textbf{\bibinfo{volume}{127}},
  \bibinfo{pages}{181802} (\bibinfo{year}{2021}), \eprint{2104.12624}.

\bibitem[{\citenamefont{Martin~Camalich
  et~al.}(2020)\citenamefont{Martin~Camalich, Pospelov, Vuong, Ziegler, and
  Zupan}}]{MartinCamalich:2020dfe}
\bibinfo{author}{\bibfnamefont{J.}~\bibnamefont{Martin~Camalich}},
  \bibinfo{author}{\bibfnamefont{M.}~\bibnamefont{Pospelov}},
  \bibinfo{author}{\bibfnamefont{P.~N.~H.} \bibnamefont{Vuong}},
  \bibinfo{author}{\bibfnamefont{R.}~\bibnamefont{Ziegler}}, \bibnamefont{and}
  \bibinfo{author}{\bibfnamefont{J.}~\bibnamefont{Zupan}},
  \bibinfo{journal}{Phys. Rev. D} \textbf{\bibinfo{volume}{102}},
  \bibinfo{pages}{015023} (\bibinfo{year}{2020}), \eprint{2002.04623}.

\bibitem[{\citenamefont{Buras et~al.}(2015)\citenamefont{Buras, Girrbach-Noe,
  Niehoff, and Straub}}]{Buras:2014fpa}
\bibinfo{author}{\bibfnamefont{A.~J.} \bibnamefont{Buras}},
  \bibinfo{author}{\bibfnamefont{J.}~\bibnamefont{Girrbach-Noe}},
  \bibinfo{author}{\bibfnamefont{C.}~\bibnamefont{Niehoff}}, \bibnamefont{and}
  \bibinfo{author}{\bibfnamefont{D.~M.} \bibnamefont{Straub}},
  \bibinfo{journal}{JHEP} \textbf{\bibinfo{volume}{02}}, \bibinfo{pages}{184}
  (\bibinfo{year}{2015}), \eprint{1409.4557}.

\bibitem[{\citenamefont{Bailey et~al.}(2016)}]{Bailey:2015dka}
\bibinfo{author}{\bibfnamefont{J.~A.} \bibnamefont{Bailey}}
  \bibnamefont{et~al.}, \bibinfo{journal}{Phys. Rev. D}
  \textbf{\bibinfo{volume}{93}}, \bibinfo{pages}{025026}
  (\bibinfo{year}{2016}), \eprint{1509.06235}.

\bibitem[{\citenamefont{Abudin\'en et~al.}(2020)}]{Belle-II:2020jti}
\bibinfo{author}{\bibfnamefont{F.}~\bibnamefont{Abudin\'en}}
  \bibnamefont{et~al.} (\bibinfo{collaboration}{Belle-II}),
  \bibinfo{journal}{Phys. Rev. Lett.} \textbf{\bibinfo{volume}{125}},
  \bibinfo{pages}{161806} (\bibinfo{year}{2020}), \eprint{2007.13071}.

\bibitem[{\citenamefont{Buras et~al.}(2021)\citenamefont{Buras, Crivellin,
  Kirk, Manzari, and Montull}}]{Buras:2021btx}
\bibinfo{author}{\bibfnamefont{A.~J.} \bibnamefont{Buras}},
  \bibinfo{author}{\bibfnamefont{A.}~\bibnamefont{Crivellin}},
  \bibinfo{author}{\bibfnamefont{F.}~\bibnamefont{Kirk}},
  \bibinfo{author}{\bibfnamefont{C.~A.} \bibnamefont{Manzari}},
  \bibnamefont{and} \bibinfo{author}{\bibfnamefont{M.}~\bibnamefont{Montull}},
  \bibinfo{journal}{JHEP} \textbf{\bibinfo{volume}{06}}, \bibinfo{pages}{068}
  (\bibinfo{year}{2021}), \eprint{2104.07680}.

\bibitem[{\citenamefont{Bennett et~al.}(2006)}]{Bennett:2006fi}
\bibinfo{author}{\bibfnamefont{G.~W.} \bibnamefont{Bennett}}
  \bibnamefont{et~al.} (\bibinfo{collaboration}{Muon g-2}),
  \bibinfo{journal}{Phys. Rev. D} \textbf{\bibinfo{volume}{73}},
  \bibinfo{pages}{072003} (\bibinfo{year}{2006}), \eprint{hep-ex/0602035}.

\bibitem[{\citenamefont{Abi et~al.}(2021)}]{Abi:2021gix}
\bibinfo{author}{\bibfnamefont{B.}~\bibnamefont{Abi}} \bibnamefont{et~al.}
  (\bibinfo{collaboration}{Muon g-2}), \bibinfo{journal}{Phys. Rev. Lett.}
  \textbf{\bibinfo{volume}{126}}, \bibinfo{pages}{141801}
  (\bibinfo{year}{2021}), \eprint{2104.03281}.

\bibitem[{\citenamefont{Aoyama et~al.}(2020)}]{Aoyama:2020ynm}
\bibinfo{author}{\bibfnamefont{T.}~\bibnamefont{Aoyama}} \bibnamefont{et~al.},
  \bibinfo{journal}{Phys. Rept.} \textbf{\bibinfo{volume}{887}},
  \bibinfo{pages}{1} (\bibinfo{year}{2020}), \eprint{2006.04822}.

\bibitem[{\citenamefont{Aoyama et~al.}(2012)\citenamefont{Aoyama, Hayakawa,
  Kinoshita, and Nio}}]{Aoyama:2012wk}
\bibinfo{author}{\bibfnamefont{T.}~\bibnamefont{Aoyama}},
  \bibinfo{author}{\bibfnamefont{M.}~\bibnamefont{Hayakawa}},
  \bibinfo{author}{\bibfnamefont{T.}~\bibnamefont{Kinoshita}},
  \bibnamefont{and} \bibinfo{author}{\bibfnamefont{M.}~\bibnamefont{Nio}},
  \bibinfo{journal}{Phys. Rev. Lett.} \textbf{\bibinfo{volume}{109}},
  \bibinfo{pages}{111808} (\bibinfo{year}{2012}), \eprint{1205.5370}.

\bibitem[{\citenamefont{Aoyama et~al.}(2019)\citenamefont{Aoyama, Kinoshita,
  and Nio}}]{Aoyama:2019ryr}
\bibinfo{author}{\bibfnamefont{T.}~\bibnamefont{Aoyama}},
  \bibinfo{author}{\bibfnamefont{T.}~\bibnamefont{Kinoshita}},
  \bibnamefont{and} \bibinfo{author}{\bibfnamefont{M.}~\bibnamefont{Nio}},
  \bibinfo{journal}{Atoms} \textbf{\bibinfo{volume}{7}}, \bibinfo{pages}{28}
  (\bibinfo{year}{2019}).

\bibitem[{\citenamefont{Czarnecki et~al.}(2003)\citenamefont{Czarnecki,
  Marciano, and Vainshtein}}]{Czarnecki:2002nt}
\bibinfo{author}{\bibfnamefont{A.}~\bibnamefont{Czarnecki}},
  \bibinfo{author}{\bibfnamefont{W.~J.} \bibnamefont{Marciano}},
  \bibnamefont{and}
  \bibinfo{author}{\bibfnamefont{A.}~\bibnamefont{Vainshtein}},
  \bibinfo{journal}{Phys. Rev. D} \textbf{\bibinfo{volume}{67}},
  \bibinfo{pages}{073006} (\bibinfo{year}{2003}), \bibinfo{note}{[Erratum:
  Phys.Rev.D 73, 119901 (2006)]}, \eprint{hep-ph/0212229}.

\bibitem[{\citenamefont{Gnendiger et~al.}(2013)\citenamefont{Gnendiger,
  St\"ockinger, and St\"ockinger-Kim}}]{Gnendiger:2013pva}
\bibinfo{author}{\bibfnamefont{C.}~\bibnamefont{Gnendiger}},
  \bibinfo{author}{\bibfnamefont{D.}~\bibnamefont{St\"ockinger}},
  \bibnamefont{and}
  \bibinfo{author}{\bibfnamefont{H.}~\bibnamefont{St\"ockinger-Kim}},
  \bibinfo{journal}{Phys. Rev. D} \textbf{\bibinfo{volume}{88}},
  \bibinfo{pages}{053005} (\bibinfo{year}{2013}), \eprint{1306.5546}.

\bibitem[{\citenamefont{Davier et~al.}(2017)\citenamefont{Davier, Hoecker,
  Malaescu, and Zhang}}]{Davier:2017zfy}
\bibinfo{author}{\bibfnamefont{M.}~\bibnamefont{Davier}},
  \bibinfo{author}{\bibfnamefont{A.}~\bibnamefont{Hoecker}},
  \bibinfo{author}{\bibfnamefont{B.}~\bibnamefont{Malaescu}}, \bibnamefont{and}
  \bibinfo{author}{\bibfnamefont{Z.}~\bibnamefont{Zhang}},
  \bibinfo{journal}{Eur. Phys. J. C} \textbf{\bibinfo{volume}{77}},
  \bibinfo{pages}{827} (\bibinfo{year}{2017}), \eprint{1706.09436}.

\bibitem[{\citenamefont{Keshavarzi et~al.}(2018)\citenamefont{Keshavarzi,
  Nomura, and Teubner}}]{Keshavarzi:2018mgv}
\bibinfo{author}{\bibfnamefont{A.}~\bibnamefont{Keshavarzi}},
  \bibinfo{author}{\bibfnamefont{D.}~\bibnamefont{Nomura}}, \bibnamefont{and}
  \bibinfo{author}{\bibfnamefont{T.}~\bibnamefont{Teubner}},
  \bibinfo{journal}{Phys. Rev. D} \textbf{\bibinfo{volume}{97}},
  \bibinfo{pages}{114025} (\bibinfo{year}{2018}), \eprint{1802.02995}.

\bibitem[{\citenamefont{Colangelo et~al.}(2019)\citenamefont{Colangelo,
  Hoferichter, and Stoffer}}]{Colangelo:2018mtw}
\bibinfo{author}{\bibfnamefont{G.}~\bibnamefont{Colangelo}},
  \bibinfo{author}{\bibfnamefont{M.}~\bibnamefont{Hoferichter}},
  \bibnamefont{and} \bibinfo{author}{\bibfnamefont{P.}~\bibnamefont{Stoffer}},
  \bibinfo{journal}{JHEP} \textbf{\bibinfo{volume}{02}}, \bibinfo{pages}{006}
  (\bibinfo{year}{2019}), \eprint{1810.00007}.

\bibitem[{\citenamefont{Hoferichter et~al.}(2019)\citenamefont{Hoferichter,
  Hoid, and Kubis}}]{Hoferichter:2019gzf}
\bibinfo{author}{\bibfnamefont{M.}~\bibnamefont{Hoferichter}},
  \bibinfo{author}{\bibfnamefont{B.-L.} \bibnamefont{Hoid}}, \bibnamefont{and}
  \bibinfo{author}{\bibfnamefont{B.}~\bibnamefont{Kubis}},
  \bibinfo{journal}{JHEP} \textbf{\bibinfo{volume}{08}}, \bibinfo{pages}{137}
  (\bibinfo{year}{2019}), \eprint{1907.01556}.

\bibitem[{\citenamefont{Davier et~al.}(2020)\citenamefont{Davier, Hoecker,
  Malaescu, and Zhang}}]{Davier:2019can}
\bibinfo{author}{\bibfnamefont{M.}~\bibnamefont{Davier}},
  \bibinfo{author}{\bibfnamefont{A.}~\bibnamefont{Hoecker}},
  \bibinfo{author}{\bibfnamefont{B.}~\bibnamefont{Malaescu}}, \bibnamefont{and}
  \bibinfo{author}{\bibfnamefont{Z.}~\bibnamefont{Zhang}},
  \bibinfo{journal}{Eur. Phys. J. C} \textbf{\bibinfo{volume}{80}},
  \bibinfo{pages}{241} (\bibinfo{year}{2020}), \bibinfo{note}{[Erratum:
  Eur.Phys.J.C 80, 410 (2020)]}, \eprint{1908.00921}.

\bibitem[{\citenamefont{Keshavarzi
  et~al.}(2020{\natexlab{a}})\citenamefont{Keshavarzi, Nomura, and
  Teubner}}]{Keshavarzi:2019abf}
\bibinfo{author}{\bibfnamefont{A.}~\bibnamefont{Keshavarzi}},
  \bibinfo{author}{\bibfnamefont{D.}~\bibnamefont{Nomura}}, \bibnamefont{and}
  \bibinfo{author}{\bibfnamefont{T.}~\bibnamefont{Teubner}},
  \bibinfo{journal}{Phys. Rev. D} \textbf{\bibinfo{volume}{101}},
  \bibinfo{pages}{014029} (\bibinfo{year}{2020}{\natexlab{a}}),
  \eprint{1911.00367}.

\bibitem[{\citenamefont{Kurz et~al.}(2014)\citenamefont{Kurz, Liu, Marquard,
  and Steinhauser}}]{Kurz:2014wya}
\bibinfo{author}{\bibfnamefont{A.}~\bibnamefont{Kurz}},
  \bibinfo{author}{\bibfnamefont{T.}~\bibnamefont{Liu}},
  \bibinfo{author}{\bibfnamefont{P.}~\bibnamefont{Marquard}}, \bibnamefont{and}
  \bibinfo{author}{\bibfnamefont{M.}~\bibnamefont{Steinhauser}},
  \bibinfo{journal}{Phys. Lett. B} \textbf{\bibinfo{volume}{734}},
  \bibinfo{pages}{144} (\bibinfo{year}{2014}), \eprint{1403.6400}.

\bibitem[{\citenamefont{Melnikov and Vainshtein}(2004)}]{Melnikov:2003xd}
\bibinfo{author}{\bibfnamefont{K.}~\bibnamefont{Melnikov}} \bibnamefont{and}
  \bibinfo{author}{\bibfnamefont{A.}~\bibnamefont{Vainshtein}},
  \bibinfo{journal}{Phys. Rev. D} \textbf{\bibinfo{volume}{70}},
  \bibinfo{pages}{113006} (\bibinfo{year}{2004}), \eprint{hep-ph/0312226}.

\bibitem[{\citenamefont{Masjuan and Sanchez-Puertas}(2017)}]{Masjuan:2017tvw}
\bibinfo{author}{\bibfnamefont{P.}~\bibnamefont{Masjuan}} \bibnamefont{and}
  \bibinfo{author}{\bibfnamefont{P.}~\bibnamefont{Sanchez-Puertas}},
  \bibinfo{journal}{Phys. Rev. D} \textbf{\bibinfo{volume}{95}},
  \bibinfo{pages}{054026} (\bibinfo{year}{2017}), \eprint{1701.05829}.

\bibitem[{\citenamefont{Colangelo et~al.}(2017)\citenamefont{Colangelo,
  Hoferichter, Procura, and Stoffer}}]{Colangelo:2017fiz}
\bibinfo{author}{\bibfnamefont{G.}~\bibnamefont{Colangelo}},
  \bibinfo{author}{\bibfnamefont{M.}~\bibnamefont{Hoferichter}},
  \bibinfo{author}{\bibfnamefont{M.}~\bibnamefont{Procura}}, \bibnamefont{and}
  \bibinfo{author}{\bibfnamefont{P.}~\bibnamefont{Stoffer}},
  \bibinfo{journal}{JHEP} \textbf{\bibinfo{volume}{04}}, \bibinfo{pages}{161}
  (\bibinfo{year}{2017}), \eprint{1702.07347}.

\bibitem[{\citenamefont{Hoferichter et~al.}(2018)\citenamefont{Hoferichter,
  Hoid, Kubis, Leupold, and Schneider}}]{Hoferichter:2018kwz}
\bibinfo{author}{\bibfnamefont{M.}~\bibnamefont{Hoferichter}},
  \bibinfo{author}{\bibfnamefont{B.-L.} \bibnamefont{Hoid}},
  \bibinfo{author}{\bibfnamefont{B.}~\bibnamefont{Kubis}},
  \bibinfo{author}{\bibfnamefont{S.}~\bibnamefont{Leupold}}, \bibnamefont{and}
  \bibinfo{author}{\bibfnamefont{S.~P.} \bibnamefont{Schneider}},
  \bibinfo{journal}{JHEP} \textbf{\bibinfo{volume}{10}}, \bibinfo{pages}{141}
  (\bibinfo{year}{2018}), \eprint{1808.04823}.

\bibitem[{\citenamefont{G\'erardin et~al.}(2019)\citenamefont{G\'erardin,
  Meyer, and Nyffeler}}]{Gerardin:2019vio}
\bibinfo{author}{\bibfnamefont{A.}~\bibnamefont{G\'erardin}},
  \bibinfo{author}{\bibfnamefont{H.~B.} \bibnamefont{Meyer}}, \bibnamefont{and}
  \bibinfo{author}{\bibfnamefont{A.}~\bibnamefont{Nyffeler}},
  \bibinfo{journal}{Phys. Rev. D} \textbf{\bibinfo{volume}{100}},
  \bibinfo{pages}{034520} (\bibinfo{year}{2019}), \eprint{1903.09471}.

\bibitem[{\citenamefont{Bijnens et~al.}(2019)\citenamefont{Bijnens,
  Hermansson-Truedsson, and Rodr\'\i{}guez-S\'anchez}}]{Bijnens:2019ghy}
\bibinfo{author}{\bibfnamefont{J.}~\bibnamefont{Bijnens}},
  \bibinfo{author}{\bibfnamefont{N.}~\bibnamefont{Hermansson-Truedsson}},
  \bibnamefont{and}
  \bibinfo{author}{\bibfnamefont{A.}~\bibnamefont{Rodr\'\i{}guez-S\'anchez}},
  \bibinfo{journal}{Phys. Lett. B} \textbf{\bibinfo{volume}{798}},
  \bibinfo{pages}{134994} (\bibinfo{year}{2019}), \eprint{1908.03331}.

\bibitem[{\citenamefont{Colangelo et~al.}(2020)\citenamefont{Colangelo,
  Hagelstein, Hoferichter, Laub, and Stoffer}}]{Colangelo:2019uex}
\bibinfo{author}{\bibfnamefont{G.}~\bibnamefont{Colangelo}},
  \bibinfo{author}{\bibfnamefont{F.}~\bibnamefont{Hagelstein}},
  \bibinfo{author}{\bibfnamefont{M.}~\bibnamefont{Hoferichter}},
  \bibinfo{author}{\bibfnamefont{L.}~\bibnamefont{Laub}}, \bibnamefont{and}
  \bibinfo{author}{\bibfnamefont{P.}~\bibnamefont{Stoffer}},
  \bibinfo{journal}{JHEP} \textbf{\bibinfo{volume}{03}}, \bibinfo{pages}{101}
  (\bibinfo{year}{2020}), \eprint{1910.13432}.

\bibitem[{\citenamefont{Blum et~al.}(2020)\citenamefont{Blum, Christ, Hayakawa,
  Izubuchi, Jin, Jung, and Lehner}}]{Blum:2019ugy}
\bibinfo{author}{\bibfnamefont{T.}~\bibnamefont{Blum}},
  \bibinfo{author}{\bibfnamefont{N.}~\bibnamefont{Christ}},
  \bibinfo{author}{\bibfnamefont{M.}~\bibnamefont{Hayakawa}},
  \bibinfo{author}{\bibfnamefont{T.}~\bibnamefont{Izubuchi}},
  \bibinfo{author}{\bibfnamefont{L.}~\bibnamefont{Jin}},
  \bibinfo{author}{\bibfnamefont{C.}~\bibnamefont{Jung}}, \bibnamefont{and}
  \bibinfo{author}{\bibfnamefont{C.}~\bibnamefont{Lehner}},
  \bibinfo{journal}{Phys. Rev. Lett.} \textbf{\bibinfo{volume}{124}},
  \bibinfo{pages}{132002} (\bibinfo{year}{2020}), \eprint{1911.08123}.

\bibitem[{\citenamefont{Colangelo et~al.}(2014)\citenamefont{Colangelo,
  Hoferichter, Nyffeler, Passera, and Stoffer}}]{Colangelo:2014qya}
\bibinfo{author}{\bibfnamefont{G.}~\bibnamefont{Colangelo}},
  \bibinfo{author}{\bibfnamefont{M.}~\bibnamefont{Hoferichter}},
  \bibinfo{author}{\bibfnamefont{A.}~\bibnamefont{Nyffeler}},
  \bibinfo{author}{\bibfnamefont{M.}~\bibnamefont{Passera}}, \bibnamefont{and}
  \bibinfo{author}{\bibfnamefont{P.}~\bibnamefont{Stoffer}},
  \bibinfo{journal}{Phys. Lett. B} \textbf{\bibinfo{volume}{735}},
  \bibinfo{pages}{90} (\bibinfo{year}{2014}), \eprint{1403.7512}.

\bibitem[{\citenamefont{Borsanyi et~al.}(2021)}]{Borsanyi:2020mff}
\bibinfo{author}{\bibfnamefont{S.}~\bibnamefont{Borsanyi}}
  \bibnamefont{et~al.}, \bibinfo{journal}{Nature}
  \textbf{\bibinfo{volume}{593}}, \bibinfo{pages}{51} (\bibinfo{year}{2021}),
  \eprint{2002.12347}.

\bibitem[{\citenamefont{Passera et~al.}(2008)\citenamefont{Passera, Marciano,
  and Sirlin}}]{Passera:2008jk}
\bibinfo{author}{\bibfnamefont{M.}~\bibnamefont{Passera}},
  \bibinfo{author}{\bibfnamefont{W.~J.} \bibnamefont{Marciano}},
  \bibnamefont{and} \bibinfo{author}{\bibfnamefont{A.}~\bibnamefont{Sirlin}},
  \bibinfo{journal}{Phys. Rev. D} \textbf{\bibinfo{volume}{78}},
  \bibinfo{pages}{013009} (\bibinfo{year}{2008}), \eprint{0804.1142}.

\bibitem[{\citenamefont{Haller et~al.}(2018)\citenamefont{Haller, Hoecker,
  Kogler, M\"onig, Peiffer, and Stelzer}}]{Haller:2018nnx}
\bibinfo{author}{\bibfnamefont{J.}~\bibnamefont{Haller}},
  \bibinfo{author}{\bibfnamefont{A.}~\bibnamefont{Hoecker}},
  \bibinfo{author}{\bibfnamefont{R.}~\bibnamefont{Kogler}},
  \bibinfo{author}{\bibfnamefont{K.}~\bibnamefont{M\"onig}},
  \bibinfo{author}{\bibfnamefont{T.}~\bibnamefont{Peiffer}}, \bibnamefont{and}
  \bibinfo{author}{\bibfnamefont{J.}~\bibnamefont{Stelzer}},
  \bibinfo{journal}{Eur. Phys. J. C} \textbf{\bibinfo{volume}{78}},
  \bibinfo{pages}{675} (\bibinfo{year}{2018}), \eprint{1803.01853}.

\bibitem[{\citenamefont{Crivellin
  et~al.}(2020{\natexlab{b}})\citenamefont{Crivellin, Hoferichter, Manzari, and
  Montull}}]{Crivellin:2020zul}
\bibinfo{author}{\bibfnamefont{A.}~\bibnamefont{Crivellin}},
  \bibinfo{author}{\bibfnamefont{M.}~\bibnamefont{Hoferichter}},
  \bibinfo{author}{\bibfnamefont{C.~A.} \bibnamefont{Manzari}},
  \bibnamefont{and} \bibinfo{author}{\bibfnamefont{M.}~\bibnamefont{Montull}},
  \bibinfo{journal}{Phys. Rev. Lett.} \textbf{\bibinfo{volume}{125}},
  \bibinfo{pages}{091801} (\bibinfo{year}{2020}{\natexlab{b}}),
  \eprint{2003.04886}.

\bibitem[{\citenamefont{Keshavarzi
  et~al.}(2020{\natexlab{b}})\citenamefont{Keshavarzi, Marciano, Passera, and
  Sirlin}}]{Keshavarzi:2020bfy}
\bibinfo{author}{\bibfnamefont{A.}~\bibnamefont{Keshavarzi}},
  \bibinfo{author}{\bibfnamefont{W.~J.} \bibnamefont{Marciano}},
  \bibinfo{author}{\bibfnamefont{M.}~\bibnamefont{Passera}}, \bibnamefont{and}
  \bibinfo{author}{\bibfnamefont{A.}~\bibnamefont{Sirlin}},
  \bibinfo{journal}{Phys. Rev. D} \textbf{\bibinfo{volume}{102}},
  \bibinfo{pages}{033002} (\bibinfo{year}{2020}{\natexlab{b}}),
  \eprint{2006.12666}.

\bibitem[{\citenamefont{Belle-II}(2020)}]{belle2projections}
\bibinfo{author}{\bibnamefont{Belle-II}} (\bibinfo{year}{2020}),
  \urlprefix\url{https://docs.belle2.org/record/2028?ln=en}.

\bibitem[{\citenamefont{Bharucha et~al.}(2016)\citenamefont{Bharucha, Straub,
  and Zwicky}}]{Bharucha:2015bzk}
\bibinfo{author}{\bibfnamefont{A.}~\bibnamefont{Bharucha}},
  \bibinfo{author}{\bibfnamefont{D.~M.} \bibnamefont{Straub}},
  \bibnamefont{and} \bibinfo{author}{\bibfnamefont{R.}~\bibnamefont{Zwicky}},
  \bibinfo{journal}{JHEP} \textbf{\bibinfo{volume}{08}}, \bibinfo{pages}{098}
  (\bibinfo{year}{2016}), \eprint{1503.05534}.

\end{thebibliography}
\end{document}